# AUTONOMOUS VEHICLE SCHEDULING AT INTERSECTIONS BASED ON PRODUCTION LINE TECHNIQUE

---

A Thesis

Presented

to the Faculty of

California State University Dominguez Hills

---

In Partial Fulfillment

of the Requirements for the Degree

Master of Science

in

Computer Science

---

by

Nasser Aloufi

Spring 2018

# ACKNOWLEDGEMENTS

I would like to thank all the professors that contributed their times and knowledges to educate me and make me a better person. I am grateful to my mother, father, brothers, and sisters who have provided me with all kinds of support in my life. Special thanks to the Department Chair Dr. Mohsen Beheshti, and to my professor Dr. Amlan Chatterjee for being professional and sharing their knowledge about my research topic. Thank you Bonnie for your sweet soul, encouragement, and making me smile ^_^.

TABLE OF CONTENTS

# LIST OF TABLES

PAGE

LIST OF FIGURES

# ABSTRACT

This thesis considers the problem of scheduling autonomous vehicles at intersections. A new system is proposed which is more efficient and could replace the recently introduced Autonomous Intersection Management (AIM) model. The proposed system is based on the production line technique. The environment of the intersection, vehicles position, speeds, and turning are specified and determined in advance. The goal of the proposed system is to eliminate vehicle’s collision and reduce the waiting time to cross the intersection. Three different patterns of traffic flow towards the intersection have been tested. The system requires less waiting time, compared to the other models, including the random case where the flow is unpredictable. The K-Nearest Neighbors (KNN) algorithm has been used to predict vehicles making a right turn at the intersection. The experimental results show there is no chance of collision inside the intersection using the proposed model; however, the system might require more space in the traffic lane for some specific traffic patterns.

# CHAPTER 1

# INTRODUCTION

There has been a steady increase in the number of vehicles on roads for the last few decades. This has led to a surge in traffic congestion that has resulted in increased delays in travel time. With the advent of technology in vehicles leading to automation, problems related to traffic congestion can be addressed using novel methods. In the near future vehicles are expected to be completely autonomous. The necessity of creating autonomous vehicles and intelligent transportation systems is more relevant than ever before. An infrastructure for such an intelligent system can be created by making vehicles interact with each other and adjusting their routes according to the traffic flow.

The development of autonomous vehicles has six stages (SAE, 2016). Stage 1 is the “no automation” phase, where a human being is responsible for all the driving tasks without involving any automation. Stage 2 is the “driver assistance” phase, where the system is responsible for the simple tasks such as steering and acceleration or deceleration. The “partial automation” is the third stage. This phase is a continuation of stage 2 with some additional features such as cruise control and lane centering. Stage 4 is the “conditional automation” phase, where the autonomous driving system performs all tasks; however, a human being should be standing by, in case anything goes wrong. Stage 5 is the “high automation” phase, where the vehicle operates autonomously in all conditions inside the domain of traffic management. Any scenario that happens in this domain, the vehicle should be able to handle it autonomously. Finally, we have the “complete automation” stage where human intervention is not needed. In this phase, the system’s performance is equivalent to that of a human being in all scenarios without any constraints.

It is expected that all vehicles will reach the full automation stage and work without any human intervention by the year 2035 (Bimbraw, 2015). The Institute of Electrical and Electronics Engineers (IEEE) claims that driverless vehicles will be the most viable form of intelligent transportation. They estimate that up to 75% of all vehicles around the world will be autonomous by the year 2040 (IEEE News Releases 2012). The sequence of the stages as described above is shown below in Figure 1.

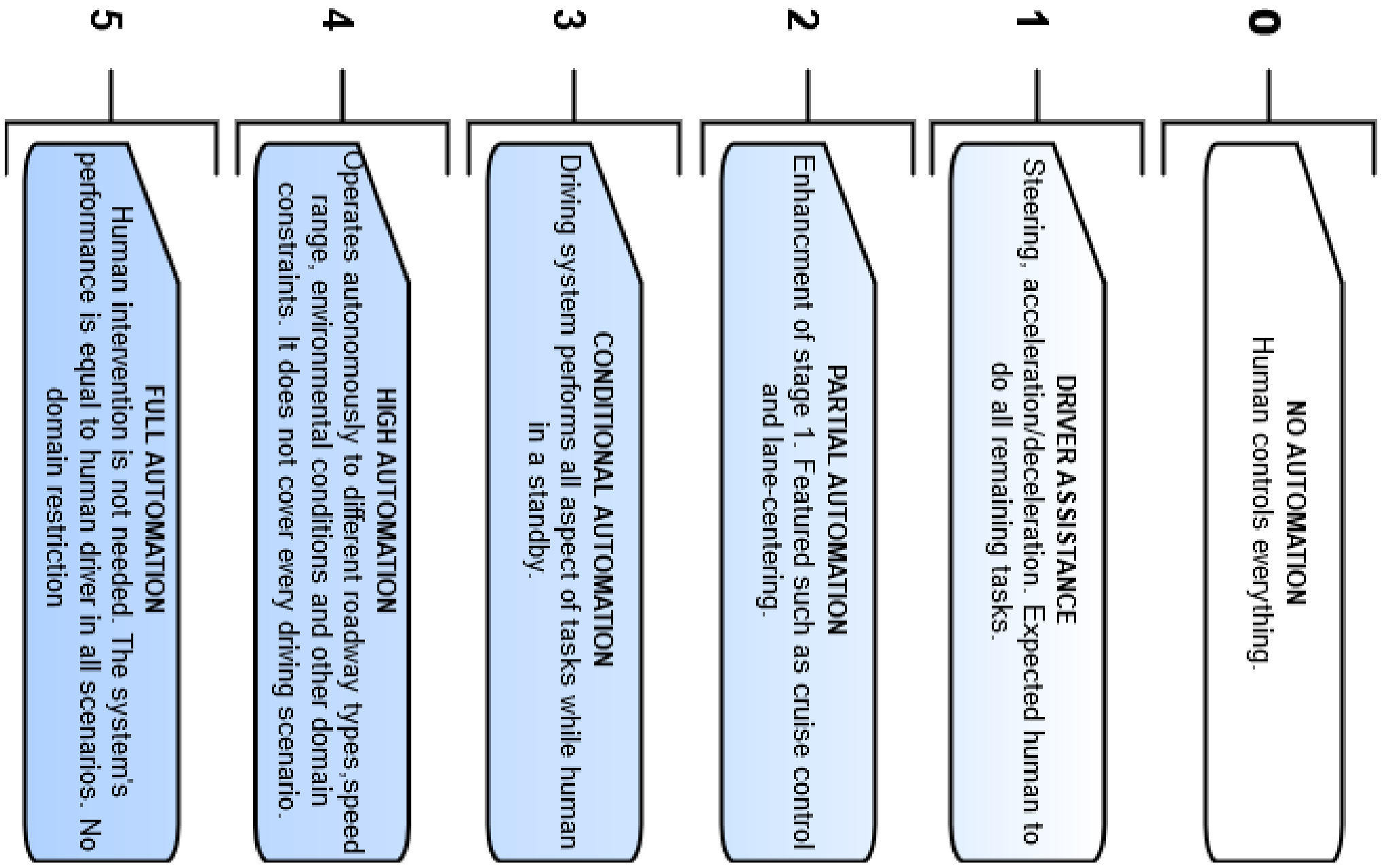

*Figure 1.* Stages of automation. Developed by thesis’s author.

In the recent years (specifically after 2008), many researches have contributed to overcome the issues of developing a scheduling system for autonomous vehicles. In the following section, we discuss some of the relevant contributions.

## Background Survey

In 2008, Defense Advanced Research Projects Agency (DARPA) organized an event which required teams to build autonomous vehicles that have the ability to drive in traffic, make maneuvers, and park. This was the first event where autonomous vehicles make interaction between manned and unmanned vehicles in a real environment (Archive.darpa.mil, 2017). VisLab in Italy contributed significant amount of research in the area of autonomous vehicles. They made numerous advance driver assistance systems (Chen et al., 2009; Cerri et al., 2010; Medici et al., 2008; Bertozzi, Broggi, & Fascioli, 2006), and created prototype vehicles such as ARGO (Bertozzi et al., 1998), TerraMax (Chen et al., 2009; Braid et al., 2006; Chen et al., 2008) and BRAiVE (Bombini et al., 2009; Grisleri & Fedriga, 2010).

In 2013, an Italian team consisting of four scientists made a long trip experiment with four autonomous vehicles (Bertozzi et al., 2013). The trip took place from Italy to China crossing more than 15,000 km for three consecutive months. The result of their experiment revealed three major challenges: (1) the autonomous vehicle had issues while making maneuvers; (2) being one of the few vehicles that follow the street rules might cause a long waiting time; (3) it is hard to combine autonomous with non-autonomous vehicles in the same environment.

An intersection algorithm model based on Mixed-Integer Linear Program (MILP) controller has been proposed (Fayazi et al., 2017). The contributions of this research were: (1) an algorithm that predicts vehicle arrivals; (2) applying mixed-integer linear program; (3) developing a simulation based on the proposed MILP controller. For maneuvering and changing lanes, a politely change lane (PCL) technique has been proposed (Hu et al., 2012). The main goal of PCL is to provide safety and efficiency while maneuvering and changing lanes.

In 2013, a self-organizing control framework for driverless vehicles was proposed based on the cooperative control framework and intersections as agent systems (Mladenovic & Abbas, 2013). The distributed intelligence was used to get the vehicle's velocity. The priority level system determines the first vehicle that passes the intersection and then adjusts the velocity of the following vehicle. The complexity of the scheduling problem has been studied using simulation based on the AIM framework (Yan et al., 2014). In this method, the problem of autonomous vehicles scheduling has been converted to a single machine scheduling problem; this fact has been used to prove that it is an NP-hard problem.

A new intersection control mechanism called Autonomous Intersection Management (AIM) was proposed (Dresner & Stone, 2008). The research results propose making a smart intersection controlling system, which would lead to vehicles' flow being more efficient than the current situation (traffic signals and stop signs). In this framework, the drivers and the designed intersections should be treated as agents. The system could have more than one agent resulting in a Multi Agent Systems (MAS).

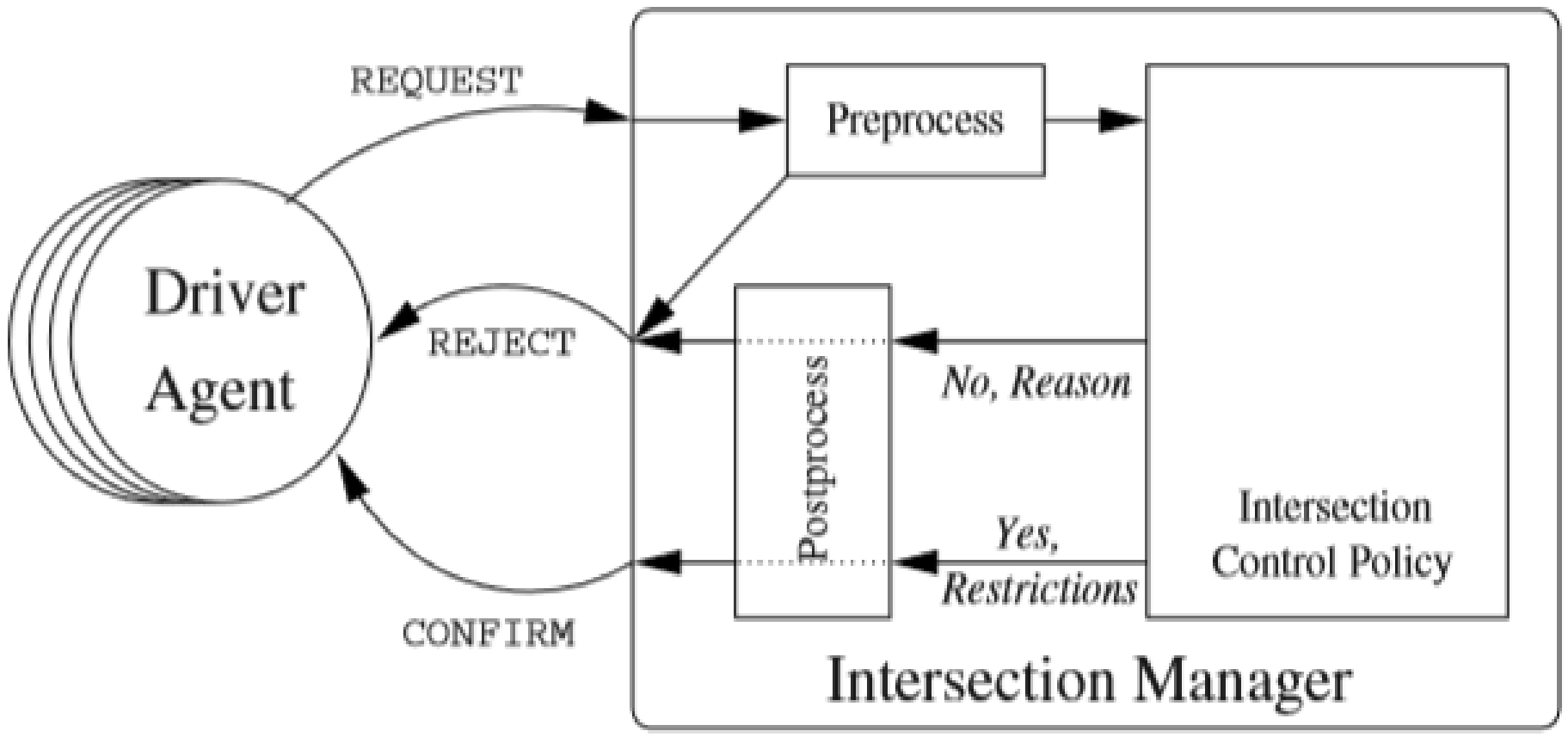

*Figure 2*. Process for AIM driver agent to make a reservation. Adapted from "A multiagent approach to autonomous intersection management" by Dresner. K, and Stone. P, 2008, *Journal of artificial intelligence research*, *31*, p. 597.

The MAS includes all the interacting elements in the environment such as drivers, pedestrians, speeds, and road signs. Whenever a vehicle wants to reserve a place in the intersection, it sends a request to the intersection manager, and the intersection manager takes an action by either accepting or rejecting the vehicle's request. Figure 2 shows the functionality of the AIM system.

# CHAPTER 2

## PROBLEM DESCRIPTION AND METHODOLOGY

It requires a complex mechanism to make all vehicles go through an intersection safely. We need to take into account the speed, timing, distance, the chance of collision, and the appropriate response needed in case of a collision. In AIM and similar systems, the vehicle sends a request to the intersection manager asking for permission to go through the intersection. The intersection manager must make one of the following decisions: (1) accept vehicle's request (when the vehicle meets the requirements); (2) reject vehicle's request (when the vehicle fails to meet all the requirements). The rejected request has two options too: (a) requires the vehicle to accelerate or decelerate (in this case the vehicle should send a new request to the intersection manager); (b) requires the vehicle to stop due to requirements failure. Figure 3 shows a general overview of the current scheduling model.

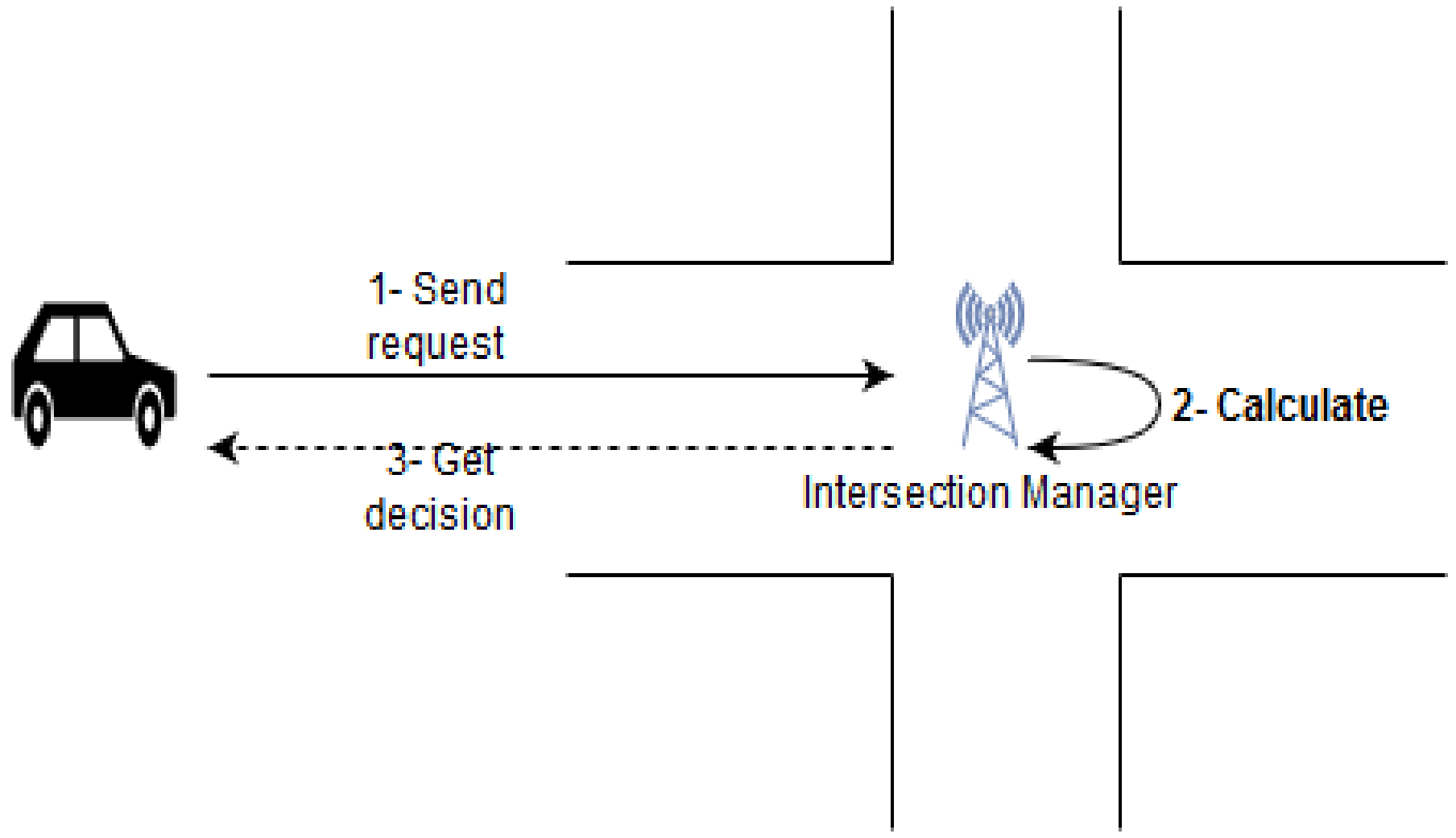

*Figure 3*. The structure of the current system. Developed by thesis's author.

It is evident that the current scheduling model has major problems related to waiting time and chance of collision. Requiring vehicles to send requests few hundred meters before approaching the intersection causes a critical time situation. There are some conditions where stopping the vehicle completely is necessary. That happens whenever $V_x(T,S,P,D) = V_z(T,S,P,D)$; where: T is the time of approaching a specific point in the intersection, S is the speed of the vehicle, P is the meeting point in the intersection, and D is the distance to that point from the current location. In addition, scheduling vehicles with a total processing time that is more than the given limited time is hard to solve. Consider we have two sets of vehicles that are going through a shared point in the intersection: Set 1 has a number of vehicles given by $V\{V_1, V_2, V_3\ldots\ldots V_n\}$, where each vehicle needs a certain amount of time to pass the shared point $M\{M_1, M_2, M_3\ldots\ldots M_n\}$. Set 2 has a number of vehicles given by $H\{H_1, H_2, H_3\ldots\ldots H_n\}$, and the respective times to pass the intersection are $S\{S_1, S_2, S_3\ldots\ldots S_n\}$. Let us assume Set 1 arrived to the intersection before Set 2, and the intersection manager assigned time constraints for each set to pass the intersection (T(X) for the first set, and T(Y) for the second set). In this case, whenever $\sum_{i=1}^{n} Mi$ > T(X), the waiting inside the intersection for T(Y) is inevitable.

In the proposed research methods, we build virtual intersections to analyze the current scheduling model. The chosen programming language for implementing the algorithms is Java. We start by calculating the waiting time and chance of collisions of the current model. Next, we propose our system and develop it. In the next stage, we test three patterns of the flow. The goal of creating these three patterns is to test different scenarios and to expand our research domain so that we can apply the proposed system to real-world intersection environment. Then, we compare the results with that of our proposed model. Figure 4 shows the general structure of our research. The key difference between the current model and our model is that the current model requires vehicle to send the request prior to making any scheduling reservation decision. However, in our system the potential location for the vehicles and the entire scheduling pattern is done before receiving the requests from vehicles to cross the intersection arrives.

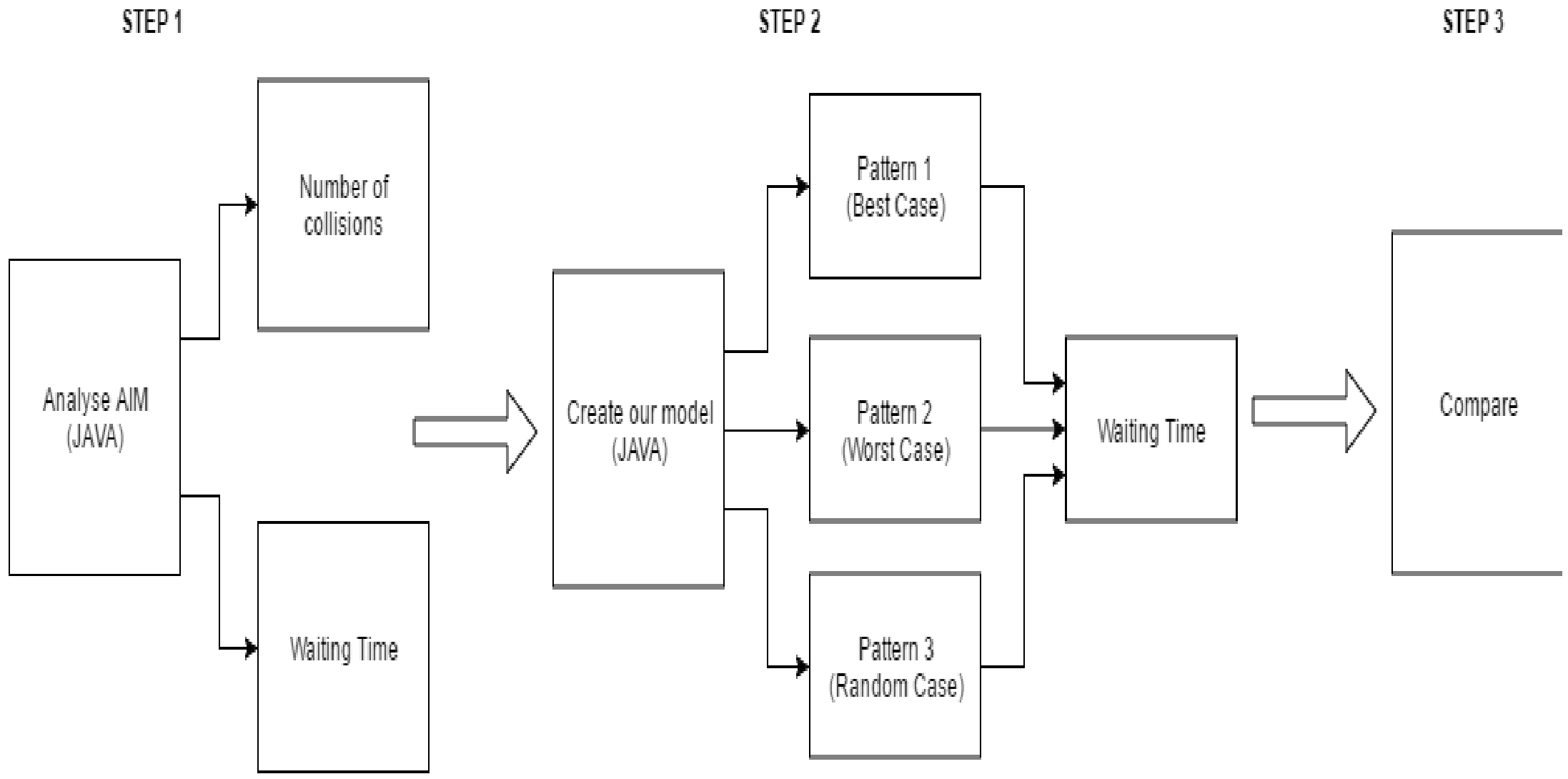

*Figure 4*. Phases of proposed research methodology. Developed by thesis's author.

## Testing Current Model

Traffic congestion is directly proportional to an increase in the case of the intersection manager requiring vehicles to stop; this is prevalent in big cities with considerable traffic and can potentially make the system inefficient. Figure 5 shows one of the conditions that causes traffic congestion. Let us consider an example and assume that we have two vehicles. Vehicle A is going to the east side with a speed of 80 mph, and vehicle B is going to the west side with the same speed. Both have the same distance to the intersection, 600 ft. Both have sent requests to the intersection manager requesting to go through the intersection. The intersection manager calculated the characteristics and the conditions of the two vehicles and came to the conclusion that both vehicles will approach the same point at the same time, specifically after 5.11 seconds. To avoid the collision, the intersection manager has two options, as we stated before: either rejects the requests and asks vehicle A or B to change its speed; or simply asks one of them to stop in order to avoid the collision.

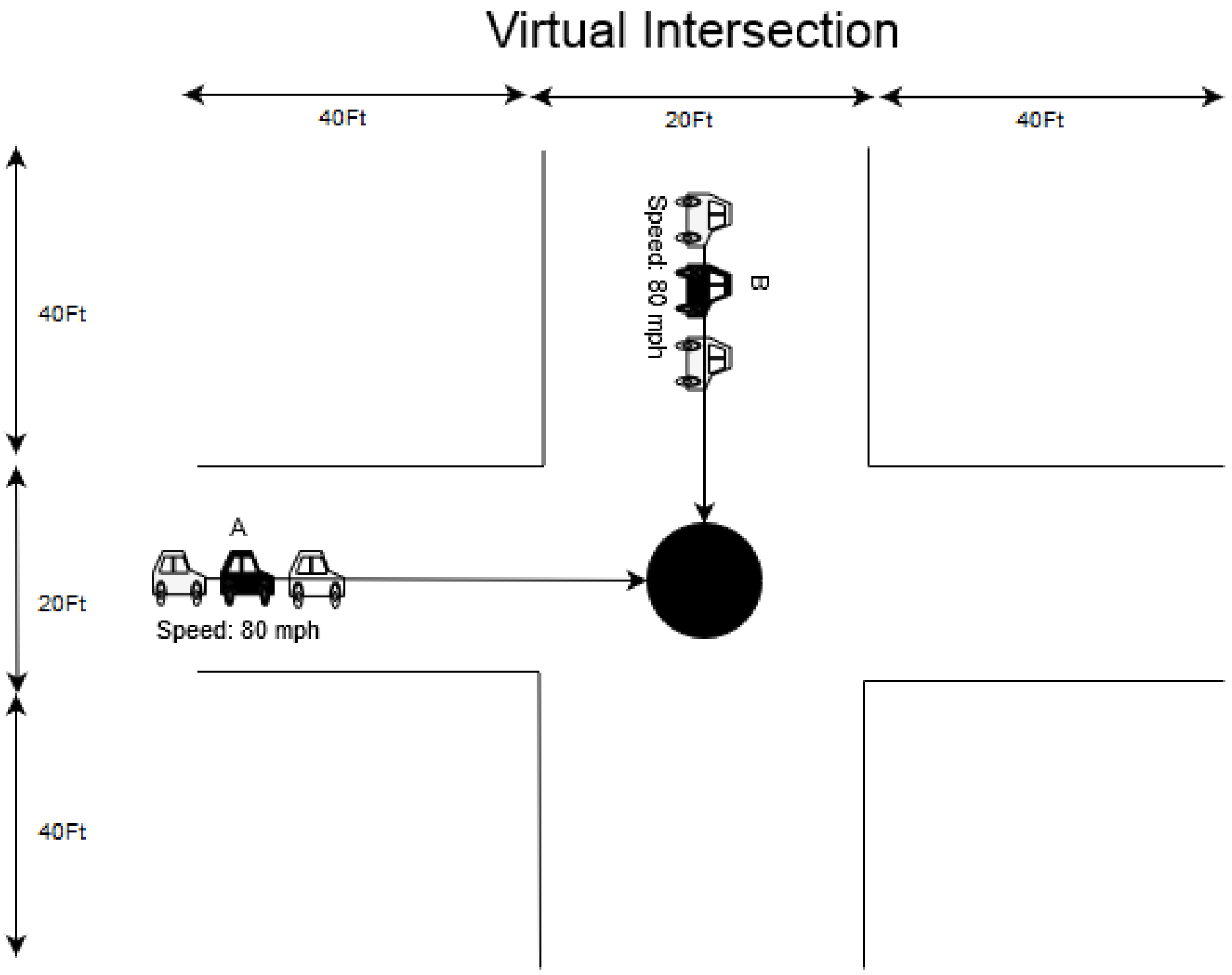

*Figure 5*. Point of collision between vehicles A and B. Developed by thesis's author.

The associated problem is that during rush hour, when the intersection gets hundreds or thousands of requests, the chances of collisions increase, leading to more stop requests and waiting times. We simulated an environment to reflect this model in order to calculate the chance of collisions that each vehicle might have. The simulated intersection has the following characteristics:

1. Two directions (One direction is going from north to south and the other one is going from west to east.).

2. Each side of the intersection can occupy 722 vehicles in total.

3. All vehicles have the same speed 100 mph.

After executing the simulation 100 times, the experimental results show that with an increase in the number of vehicles, the number of expected collisions and waiting times also increase. With 50 vehicles in the intersection, the number of expected collisions is shown to be one for each vehicle. The number of collisions go up to three when we consider 200 vehicles. With 300 vehicles, the number of collisions increased to 4.3 for each vehicle. Similar outcome is also observed for the waiting time; the waiting time increases with the number of vehicles. The waiting time is calculated to be 85 seconds per vehicle when we consider 50 vehicles inside the intersection. Then it increased to 320 seconds per vehicle when we consider 200 vehicles. Finally, with 300 vehicles, the waiting time is 515 seconds per vehicle. Figure 6 and Figure 7 show the number of collisions and waiting times respectively.

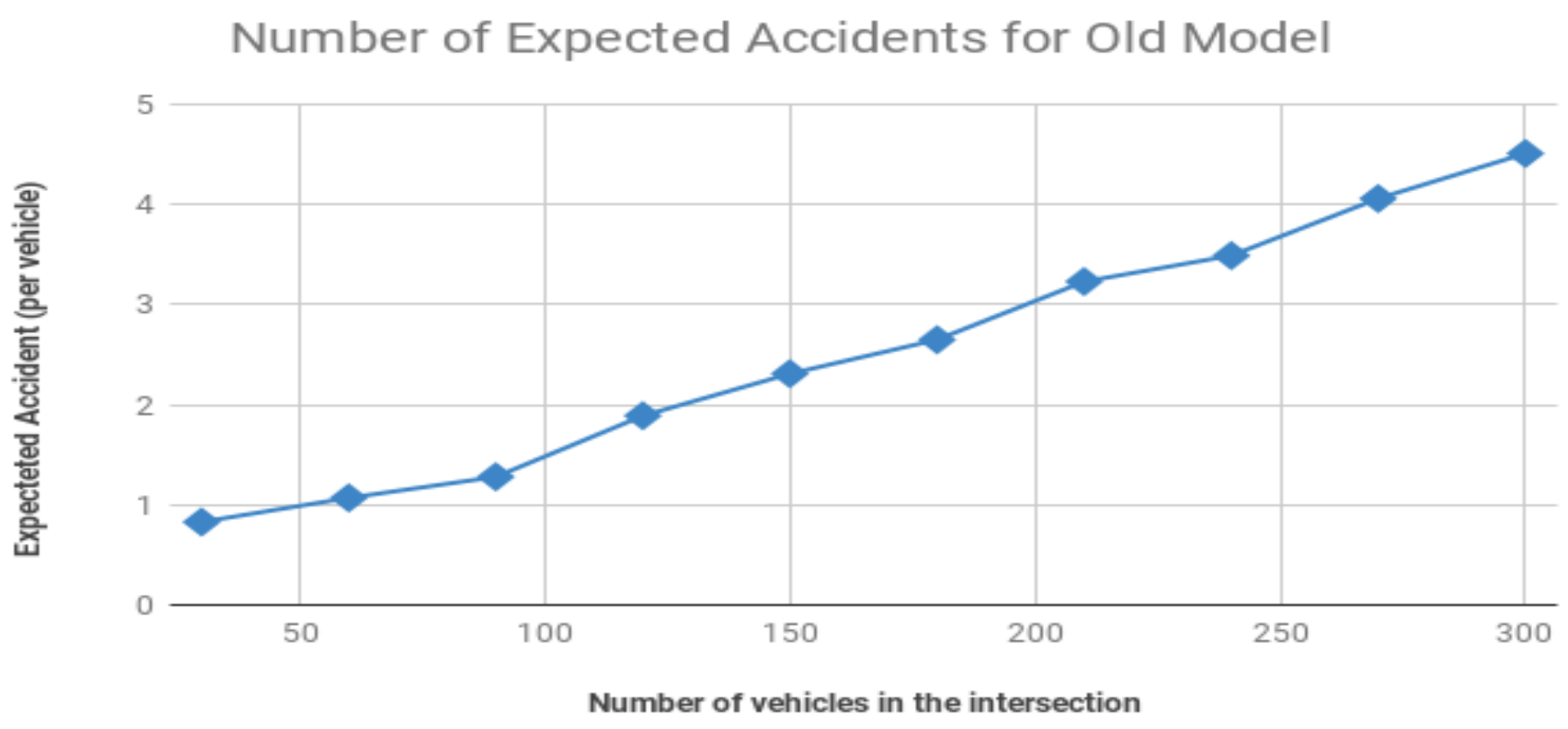

*Figure 6*. Expected number of potential collisions for every vehicle. Developed by thesis's author.

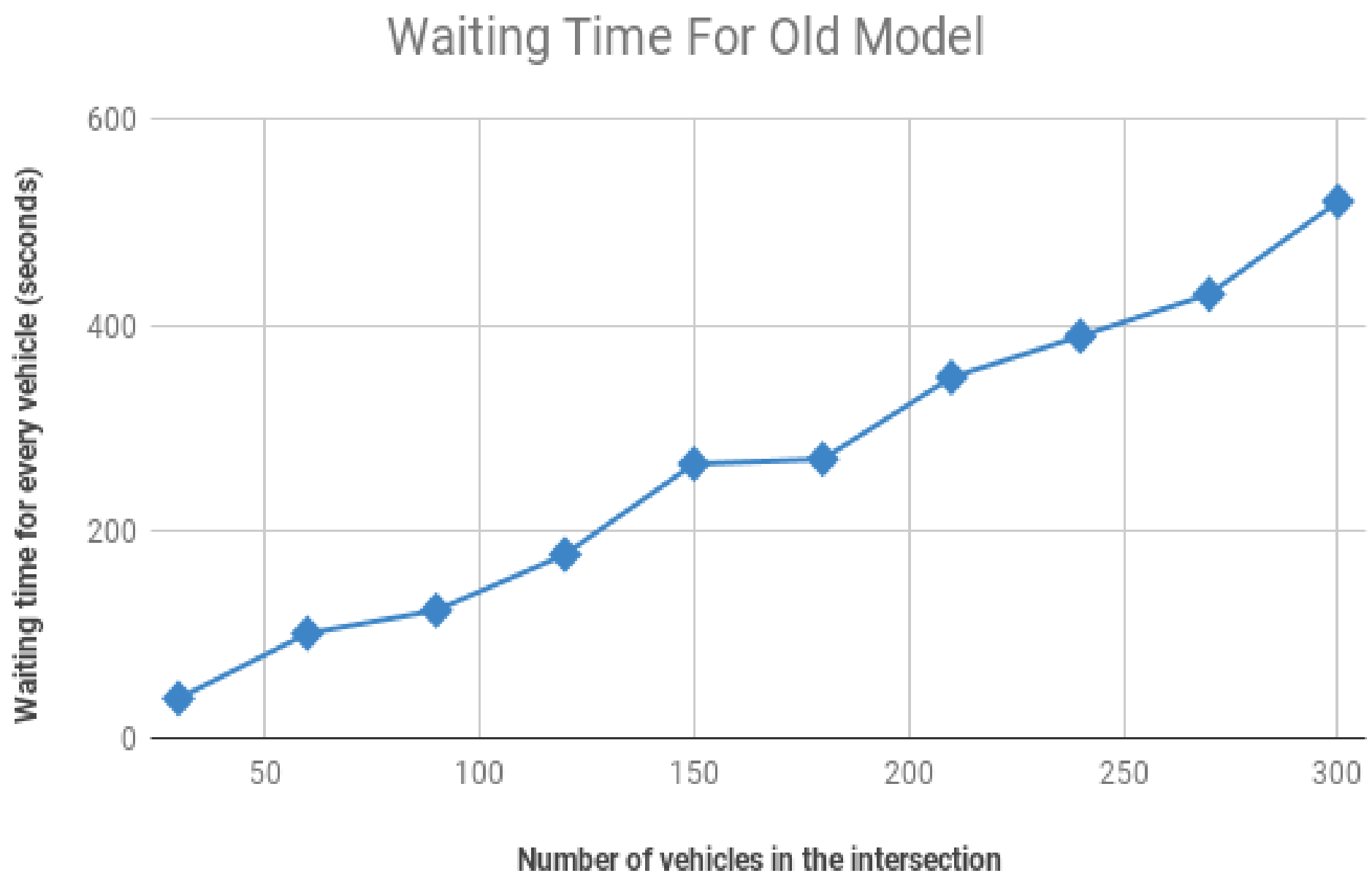

*Figure 7*. The waiting time for every vehicle. Developed by thesis's author.

Increasing and decreasing the speed is not an option when the intersection gets enormous requests because there are other vehicles in the vicinity. With vehicles in front and behind of a vehicle under consideration to be requested to change the speed, fixing the problem in this manner would lead to problems and potential collisions with the other vehicles. In summary, with an increase in the number of vehicles approaching the intersection, the chances of collision also increase, thereby leading to more stopping time.

# CHAPTER 3

# THE PROPOSED SYSTEM

In the old model, the scheduling process starts after receiving the vehicle's request (Figure 3). In the proposed method, we flipped the current approach and created another model where the scheduling should be set up in the intersection prior to receiving any request. Our technique is based on the production line system where every position (or container) in the line is reserved for a specific item.

Figure 8 shows the architecture of our system where the intersection's spots should be fixed as the first step. We prepare the intersection by making containers that are based on the length of the vehicles. Once we fix the vehicles' position, enter timing, and the speed of the lanes, then vehicles can send requests to the intersection manager. Figure 9 shows the complete design; where $S_1$ is the minimum accepted speed, $S_2$ the maximum accepted speed, "spin" is the timing where the lane is open.

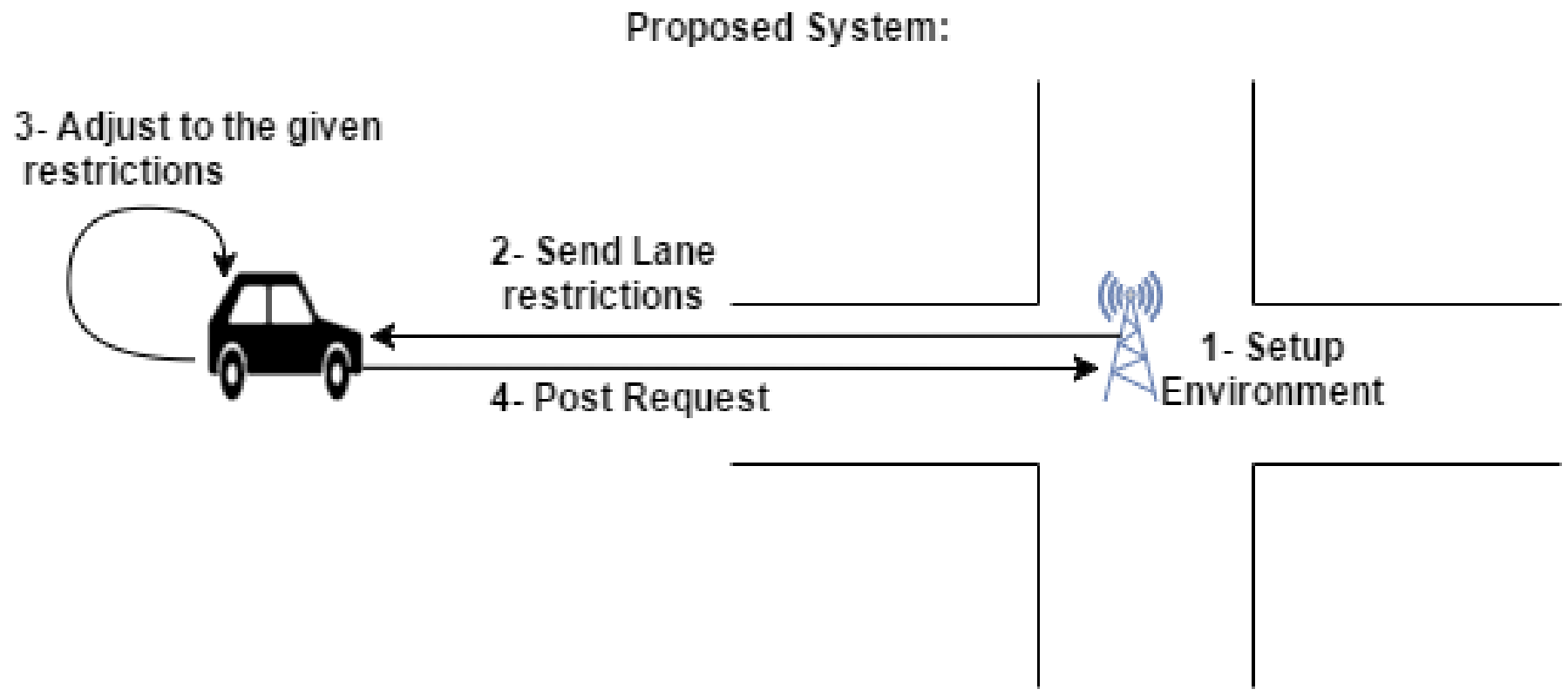

*Figure 8*. The architecture of the proposed system. Developed by thesis's author.

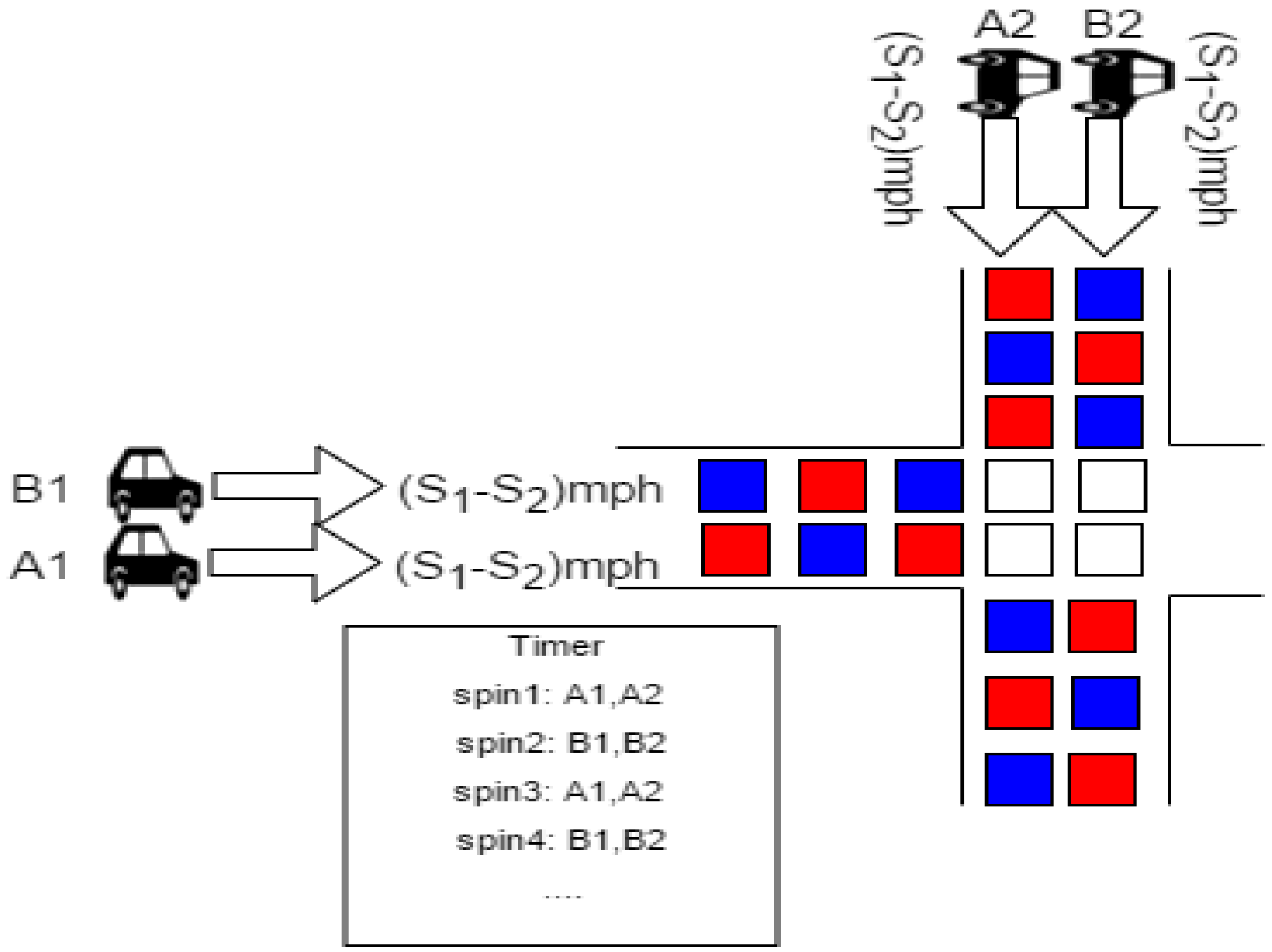

*Figure 9.* Production line intersection. Developed by thesis's author.

A timer is used to switch between lanes. If lane A is open at any moment, lane B should be closed at that instant. A setpoint system for generating setpoints for the Proportional Integral Derivative (PID) controllers was proposed (Au, Quinlan, & Stone, 2012). The purpose of the system is to make sure that the vehicle arrives at the exact expected time with the exact expected speed. However, the system doesn't guarantee the arrival time of the vehicle to a specific point, so it's not ideal where an unexpected latency can cause a catastrophic result. We took that into consideration, and instead of applying one specific speed value, we applied $S_1$ and $S_2$ which are the minimum and maximum speed. Any vehicle that has this range of speed should be accepted. Figure 10 shows the algorithm of our system.

**Algorithm 1** The proposed algorithm for production line scheduling

```
initialize lane for a certain distance
I_t ←the intersection running time
O_t ←lane's opining time
S_min ←lane's minimum speed
S_max ←lane's maximum speed
P_length ←length of the spot
V_s ←vehicle speed
V_ar ←vehicle arriving time
if V_s ≤ S_max AND V_s ≥ S_min then
    if V_ar = Lane's[O_t] then
        Set Average Speed
        Set Allow Entering
    else REJECT
else REJECT
```

*Figure 10*. Algorithm of our system. Developed by thesis's author.

Related Problems and Solutions

There are three problems that can occur using the initial design. The first potential problem can occur because of the speed variation. After applying $S_1$ and $S_2$, vehicles speed can vary, thereby leading vehicles to collide after a certain distance. Let us assume we set up the entering speed for lane A to 60 mph as a minimum speed ($S_1$), and to 65 mph as a maximum speed ($S_2$). Two vehicles ($V_1$ and $V_2$) sent requests to enter the intersection. $V_1$ came first with a 61 mph speed, and $V_2$ came after it with a 64 mph speed. The problem is that after a certain distance, $V_2$ will approach $V_1$ and collide with it.

The second problem can happen when the vehicle enters an intersection with one speed range while the following intersection requires a different speed range. Let us assume we have two intersections, one after another. The first intersection accepts vehicles with any speed between 60–65 mph, and the second one accepts vehicles with 102.5–107.5 mph as a speed range. The problem now is to make those two intersections compatible so that they become connected into one system. Figure 11. Shows this problem.

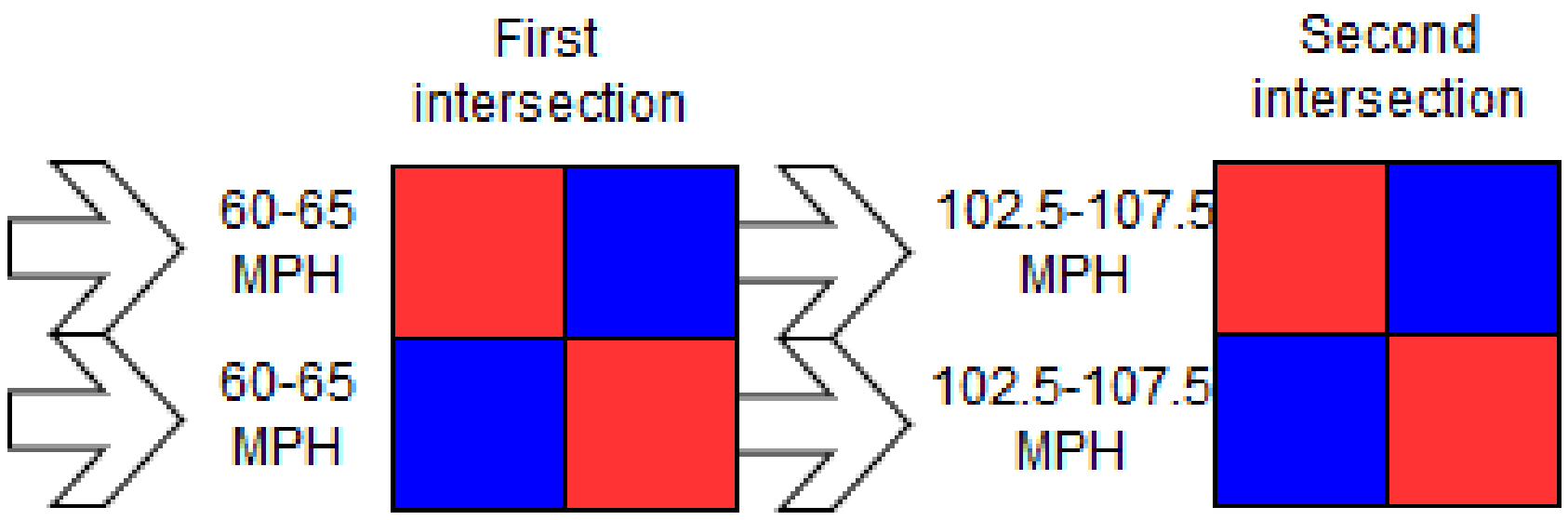

*Figure 11.* Varying speed between two intersections. Developed by thesis's author.

The third potential problem is regarding the ability of a vehicle to make a right turn. The system must produce a container that makes a right turn, and it should do it safely to avoid collision with other vehicles.

To solve the first problem, we applied an average speed value. This average speed should be applied to every vehicle as soon as it enters the intersection (Avg = $(S_1+S_2)/2$). The goal of the average speed is to ensure that each vehicle stays in its given container without jumping to another one. Figure 12 shows the problem before and after applying the average speed.

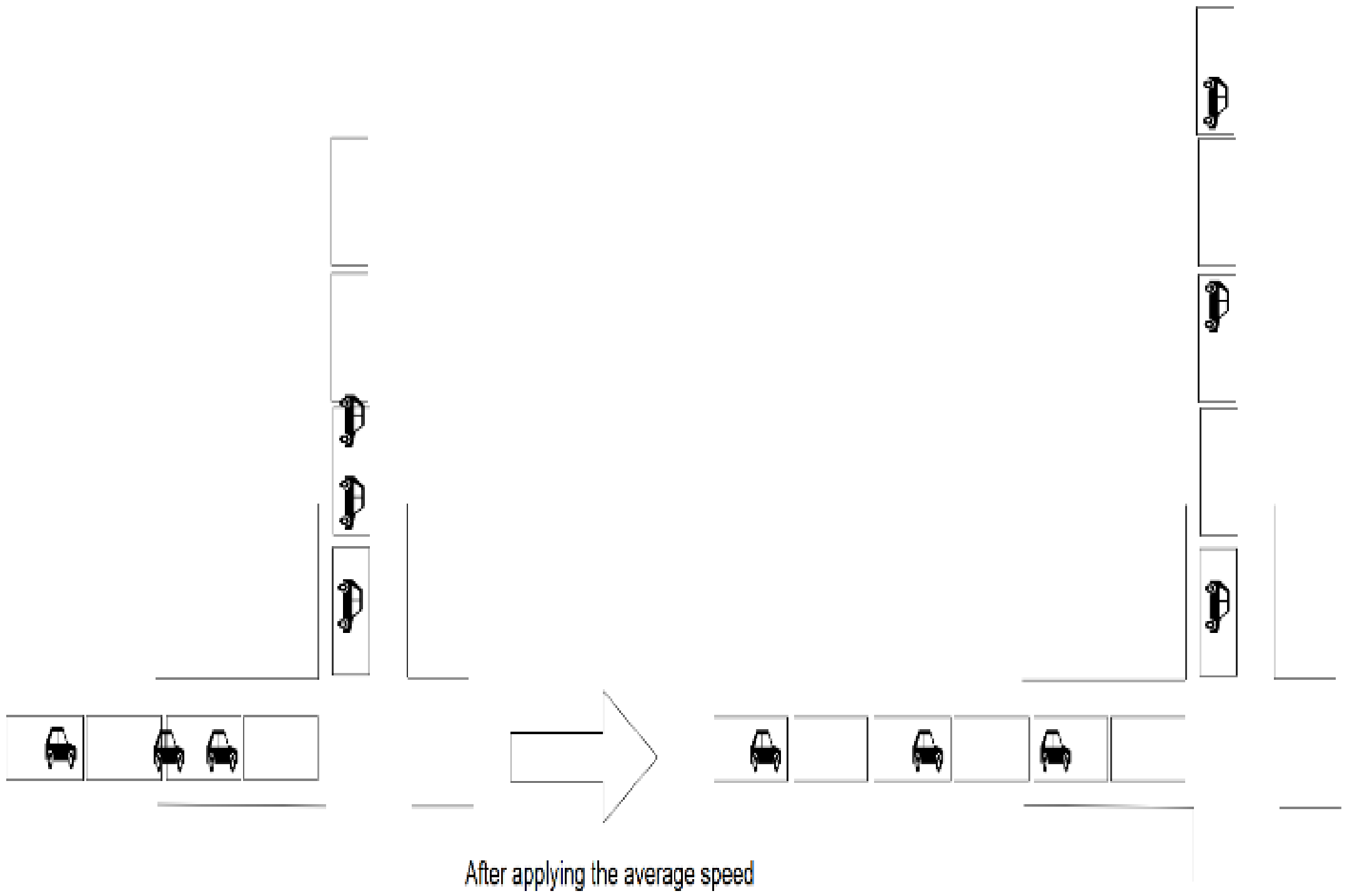

*Figure 12*. The intersection flow before and after applying the average speed. Developed by thesis's author.

To solve the second problem, we created increasing and decreasing speed areas. The intersection manager asks all the vehicles to either increase or decrease the speed depending on the speed of the next intersection. $S = Ep_1 + (Tp_2 - Ep_1)$; where: Ep is the exit speed of the intersection. Tp is the entering speed of the following intersection. Figure 13 shows how the increasing speed works.

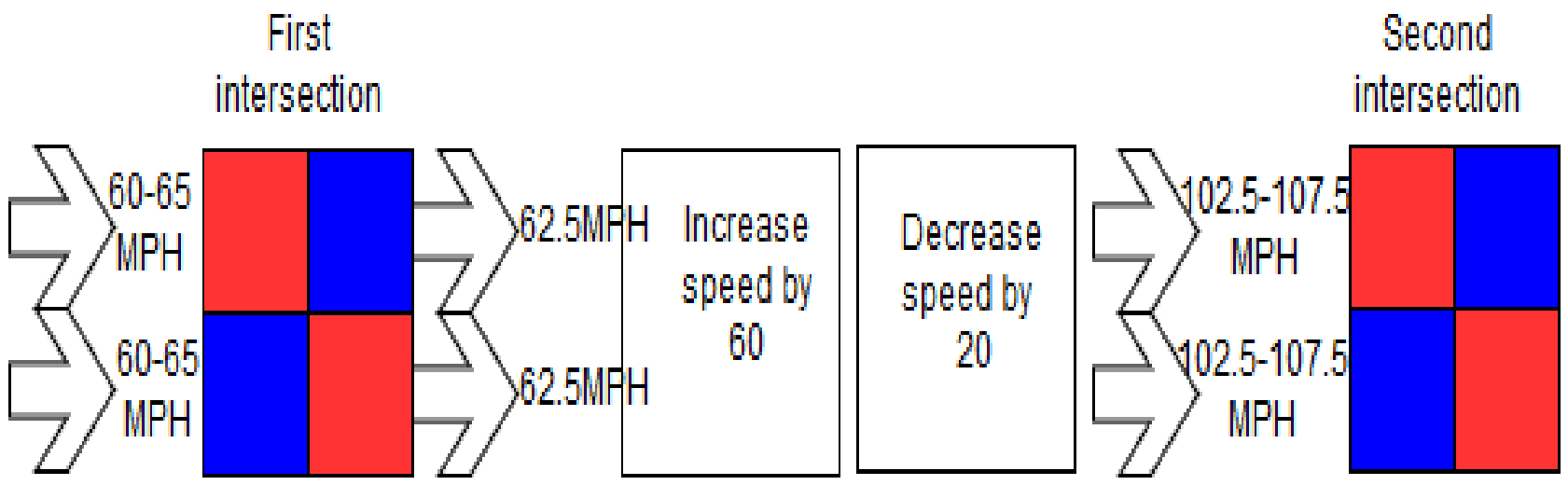

*Figure 13*. Areas of changing the speed. Developed by thesis's author.

Regarding the third problem, where we need to handle vehicles making a right turn, we designed a classifier using the K-nearest neighbor (KNN) technique. Using the classifier, the intersection can produce right turn containers based on specific features we would like to apply. Some of the features that we could apply for the KNN are: day, time, population of the city, event happening in the area, and so on. For our KNN we used the Euclidean distance function:

$$\sqrt{(q_1 - p_1)^2 + (q_2 - p_2)^2 + \cdots + (q_n - p_n)^2}$$

The classifier is used to calculate the possibility of making a right turn for the next container (vehicle). Figure 14 shows an example of the KNN method.

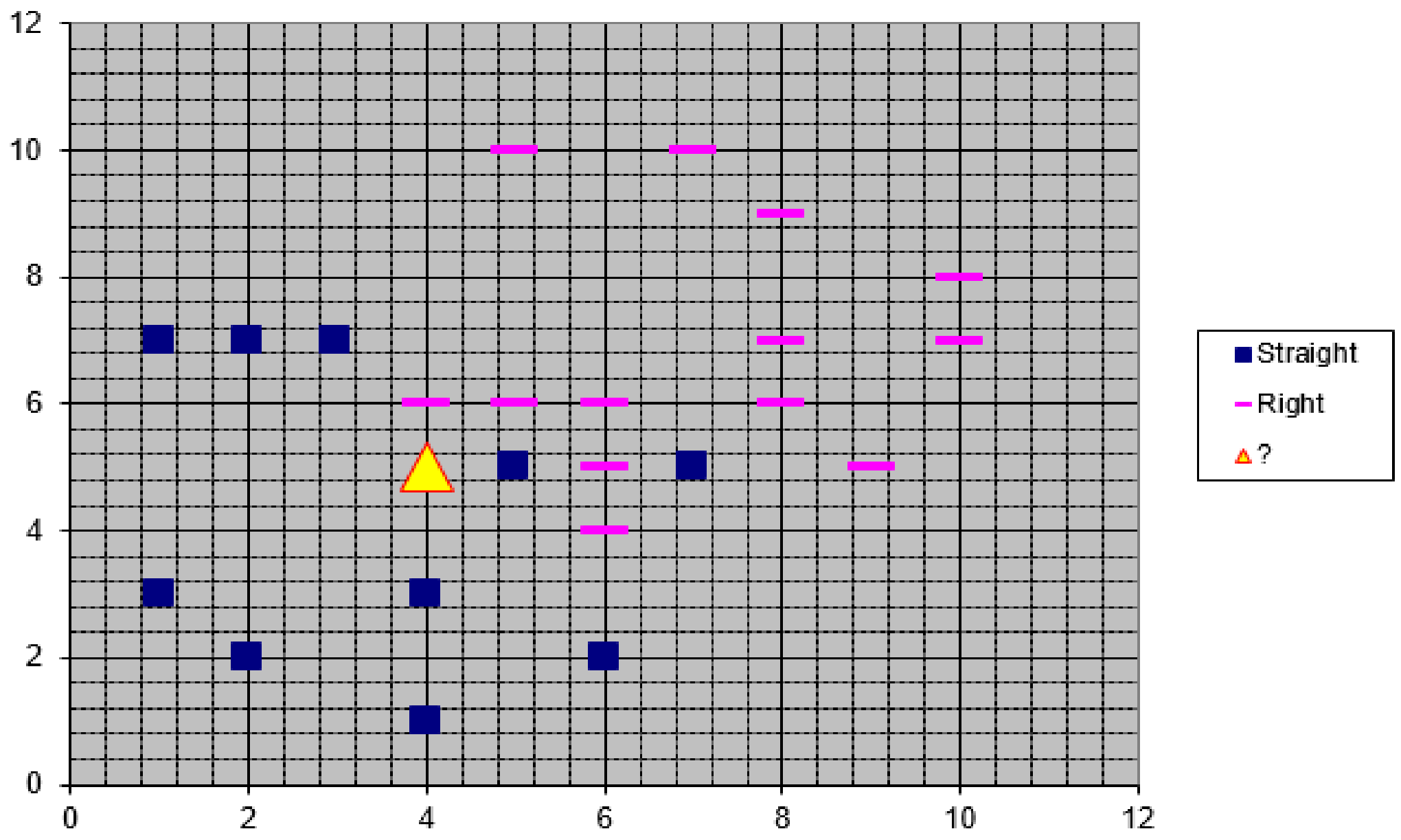

*Figure 14*. The vehicle in the yellow triangle is going to make a right turn when K=3. Developed by thesis's author.

## Simulation and Result

To evaluate our proposed techniques, we made a virtual intersection with the following features:

1. Vehicles are coming from four different directions:

   A1: from east to west.

   A2: from west to east.

   B1: from north to west.

   B2: from west to north.

2. The speeds of the arriving vehicles ranged between 60–65 mph.
3. The length of every container is 26.2467 ft.
4. The running time for the intersection is one minute.
5. The system produces one container for every opening lane.
6. Gates A1 and A2 should open together while gates B1 and B2 closed.
7. Gates B1 and B2 should open together while gates A1 and A2 closed.
8. Every lane has 60 containers in total.

Figure 15 shows an illustration of the virtual intersection that we built for the experiment. For predicting the expected right turn in the KNN we need to create initial testing data (Table 1) in order to start training the system.

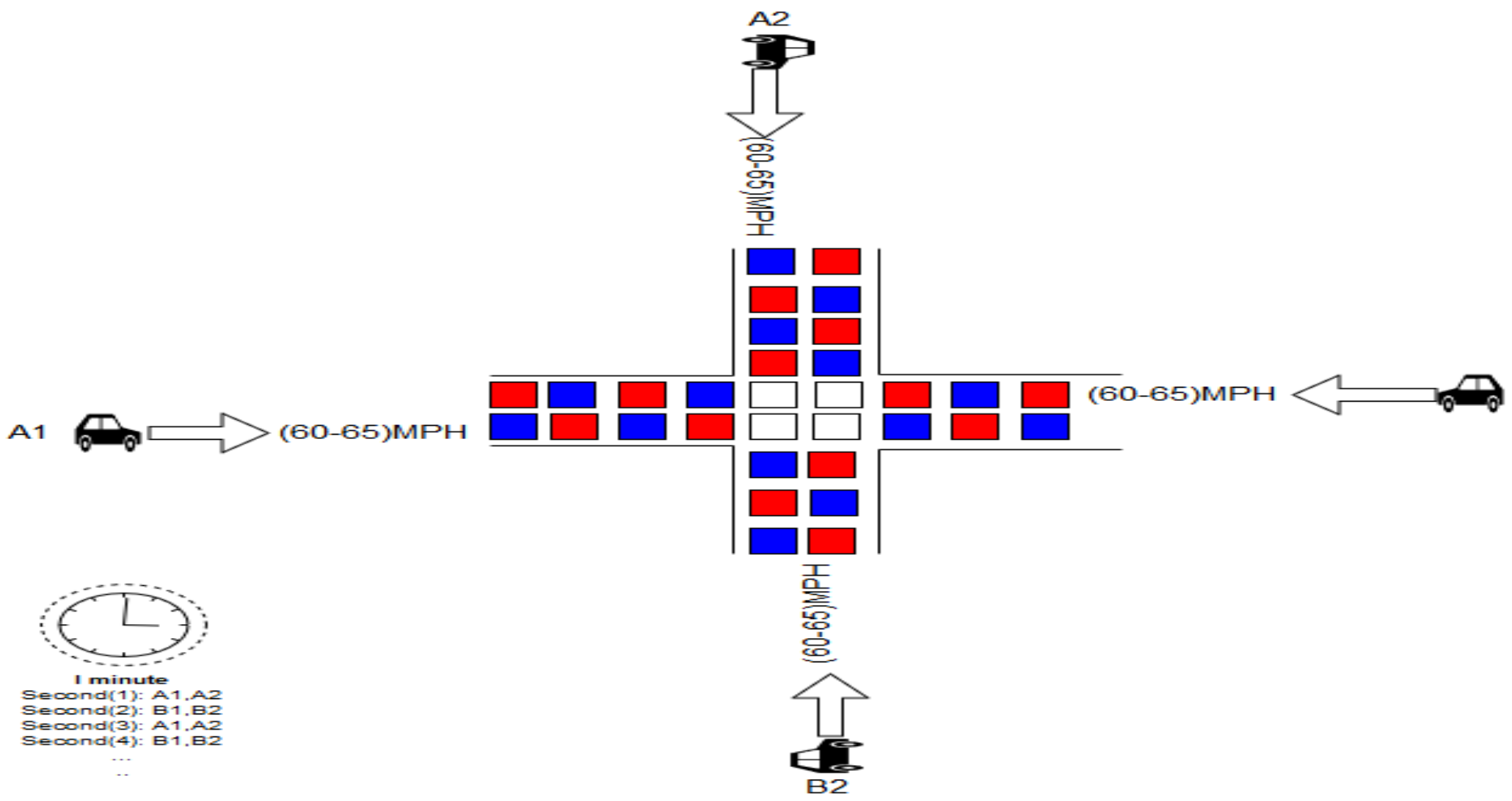

*Figure 15*. Experiment virtual intersection. Developed by thesis's author.

Table 1

*Features and Initial Data*

| Day | Hour | Event in the area | Class of turn |
|---|---|---|---|
| 1 | 9 | 0 | + |
| 3 | 10 | 0 | + |
| 4 | 8 | 0 | + |
| 3 | 8 | 0 | + |
| 4 | 10 | 0 | + |
| 2 | 20 | 1 | - |
| 5 | 19 | 1 | - |
| 1 | 4 | 1 | - |
| 2 | 7 | 1 | - |

*Note*. Developed by thesis's author.

In Table 1, days are numbered from 1 (Monday) to 5 (Friday). Hours use a 24 format. Event in the area is "0" if there is an event in the area of the intersection and "1" if there is no event. The "+" means the vehicle is going to make a right turn, while the "-" means the vehicle is moving in a straight direction.

We also created three different patterns for the flow of the coming vehicles. There are two goals of these three patterns. First, we want to calculate the waiting time for every flow. Second, we want to calculate the required space for each pattern. The first pattern represents the best case where the flow matches the opening and the closing times of the intersection. The second pattern represents the worst case of the flow where vehicles keep coming even if the intersection is closed. The third and final pattern represents the random (normal) case where the flow is unpredictable. Each of the three patterns has the follow features: (1) it has 60 spots; (2) accepts only one request per spot; (3) runs for one minute. For the patterns, we got different results based on the flow toward the intersection. For all of the results below: "1" represents a request, "0" represents no request.

In the best case, the total waiting time is always zero. This is simply because the pattern of the flow matches the intersection's requirements. The opening time matches the requests in all cases. The flow pattern in this case is: [1, 0, 1, 0, 1, 0, 1, 0, 1, 0, 1, 0, 1, 0, 1, 0, 1, 0, 1, 0, 1, 0, 1, 0, 1, 0, 1, 0, 1, 0, 1, 0, 1, 0, 1, 0, 1, 0, 1, 0, 1, 0, 1, 0, 1, 0, 1, 0, 1, 0, 1, 0, 1, 0, 1, 0, 1, 0, 1, 0]. We don't need to arrange the flow in such case; hence the arriving times in one minute run: [0, 2, 4, 6, 8, 10, 12, 14, 16, 18, 20, 22, 24, 26, 28, 30, 32, 34, 36, 38, 40, 42, 44, 46, 48, 50, 52, 54, 56, 58].

In the second case the flow is constant (worst-case): [1, 1, 1, 1, 1, 1, 1, 1, 1, 1, 1, 1, 1, 1, 1, 1 ,1 ,1 ,1 ,1 ,1 ,1 ,1 ,1 ,1 ,1 ,1 ,1 ,1 ,1, 1, 1, 1, 1, 1, 1, 1, 1, 1, 1, 1, 1, 1, 1, 1, 1, 1, 1, 1, 1, 1, 1, 1, 1, 1, 1, 1, 1, 1]. The arriving times for this pattern in one minute run: [0, 1, 2, 3, 4, 5, 6, 7, 8, 9, 10, 11, 12, 13, 14, 15, 16, 17, 18, 19, 20, 21, 22, 23, 24, 25, 26, 27, 28, 29, 30, 31, 32, 33, 34, 35, 36, 37, 38, 39, 40, 41, 42, 43, 44, 45, 46, 47, 48, 49, 50, 51, 52, 53, 54, 55, 56, 57, 58, 59]. In this pattern, arranging arriving times to match the intersection's requirements is time consuming. The average waiting time is about 164.84 seconds per vehicle, and the arranged flow in one minute is: [0, 2, 4, 6, 8, 10, 12, 14, 16, 18, 20, 22, 24, 26, 28, 30, 32, 34, 36, 38, 40, 42, 44, 46, 48, 50, 52, 54, 56, 58, 60, 62, 64, 66, 68, 70, 72, 74, 76, 78, 80, 82, 84, 86, 88, 90, 92, 94, 96, 98, 100, 102, 104, 106, 108, 110, 112, 114, 116, 118]. The major issue with this constant pattern is not only the high waiting time, but also we need to increase the space by 100% to arrange the vehicles.

The last pattern represents the random case where the flow is unpredictable. The proposed system results are very promising since the average waiting time is less than the old models. In our model, the total waiting time for the random case is only 35 seconds compared to 101 seconds in the old model. A one minute random set for such pattern would be: [1, 1, 0, 1, 1, 1, 0, 0, 0, 1, 0, 1, 0, 1, 0, 1, 0, 1, 1, 0, 0, 1, 0, 0, 1, 0, 1, 0, 1, 0, 1, 0, 1, 1, 1, 0, 1, 0, 1, 0, 0, 1, 0, 1, 0, 0, 0, 0, 1, 0, 1, 0, 1, 1, 0, 0, 0, 1, 1, 1], and the arriving times for the corresponding set: [0, 1, 3, 4, 5, 9, 11, 13, 15, 17, 18, 21, 24, 26, 28, 30, 32, 33, 34, 36, 38, 41, 43, 48, 50, 52, 53, 57, 58, 59]. The arranged flow for this set would be: [0, 0, 1, 0, 3, 0, 4, 0, 5, 0, 9, 0, 11, 0, 13, 0, 15, 0, 17, 0, 18, 0, 21, 0, 24, 0, 26, 0, 28, 0,

30, 0, 32, 0, 33, 0, 34, 0, 36, 0, 38, 0, 41, 0, 43, 0, 48, 0, 50, 0, 52, 0, 53, 0, 57, 0, 58, 0, 59]. The average waiting time for this set is only 6.76 seconds per vehicle.

To sum up results, the first pattern has 0 waiting time and 0 additional space. In the second pattern, the time and space increase by 100%. In the third pattern (random flow), the result of the waiting time depends on the number of requests: (a) 0 if the number of requests is less than or equal to the giving spots; (b) ((n-30) /30*100) if the number of requests is more than the given spots. The "n" in the previous equation is the number of requests, and 30 indicates the number of spots available to be occupied.

For predicting the right turn, whenever a new vehicle comes to the lane, the KNN classifier calculates the features (day, hour, event) and predicts whether or not the vehicle makes a right turn. The resulting data of the entered vehicle will be added to the classifier so it helps to predict the next generated spot. Table 2 and Table 3 show the instances for the four lanes, while Table 4 and Table 5 show the right turn predictions.

Table 2

*Lane A1 and A2 Instances*

| Day | Hour | Event in the area |
|---|---|---|
| 3 | 3 | 0 |
| 1 | 5 | 0 |
| 2 | 11 | 0 |
| 5 | 10 | 0 |
| 4 | 1 | 0 |

*Table Continued*

| Day | Hour | Event in the area |
|---|---|---|
| 1 | 23 | 0 |
| 2 | 13 | 0 |
| 2 | 18 | 1 |
| 3 | 14 | 1 |
| 2 | 8 | 1 |
| 4 | 6 | 1 |
| 1 | 21 | 1 |
| 2 | 3 | 0 |
| 4 | 15 | 0 |
| 4 | 22 | 1 |
| 4 | 22 | 0 |
| 3 | 8 | 1 |
| 2 | 21 | 0 |
| 3 | 0 | 1 |
| 1 | 20 | 0 |
| 4 | 12 | 0 |
| 1 | 3 | 1 |
| 1 | 7 | 1 |
| 1 | 23 | 1 |
| 2 | 2 | 1 |

*Table Continued*

| Day | Hour | Event in the area |
|---|---|---|
| 4 | 6 | 0 |
| 2 | 0 | 1 |
| 3 | 16 | 0 |
| 2 | 1 | 1 |
| 4 | 11 | 0 |

*Note.* Developed by thesis's author.

Table 3

*Lane B1 and B2 Instances*

| Day | Hour | Event in the area |
|---|---|---|
| 1 | 8 | 0 |
| 4 | 12 | 0 |
| 5 | 2 | 1 |
| 4 | 3 | 1 |
| 4 | 12 | 0 |
| 2 | 1 | 0 |
| 2 | 13 | 0 |

*Table Continued*

| Day | Hour | Event in the area |
|---|---|---|
| 4 | 20 | 1 |
| 3 | 18 | 1 |
| 5 | 16 | 0 |
| 2 | 1 | 0 |
| 5 | 3 | 0 |
| 5 | 9 | 0 |
| 3 | 11 | 1 |
| 4 | 2 | 1 |
| 2 | 9 | 0 |
| 2 | 9 | 0 |
| 5 | 19 | 1 |
| 3 | 11 | 1 |
| 3 | 18 | 1 |
| 2 | 16 | 1 |
| 5 | 15 | 1 |
| 3 | 11 | 1 |
| 4 | 17 | 1 |
| 3 | 16 | 1 |
| 3 | 22 | 1 |
| 5 | 23 | 0 |

*Table Continued*

| Day | Hour | Event in the area |
|---|---|---|
| 2 | 12 | 1 |
| 2 | 11 | 0 |
| 5 | 19 | 1 |

*Note.* Developed by thesis's author.

Table 4

*Turn Predictions For Lane A1 And Lane A2*

| 1 | 2 | 3 | 4 | 5 | 6 | 7 | 8 | 9 | 10 |
|---|---|---|---|---|---|---|---|---|---|
| - | - | + | + | - | - | + | - | + | + |
| 11 | 12 | 13 | 14 | 15 | 16 | 17 | 18 | 19 | 20 |
| + | - | - | + | - | - | + | - | - | - |
| 21 | 22 | 23 | 24 | 25 | 26 | 27 | 28 | 29 | 30 |
| + | - | + | - | - | + | - | + | - | + |

*Note.* Developed by thesis's author.

Table 5

*Turn Predictions For Lane B1 And Lane B2*

| | | | | | | | | | |
|---|---|---|---|---|---|---|---|---|---|
| 1 | 2 | 3 | 4 | 5 | 6 | 7 | 8 | 9 | 10 |
| + | + | - | - | + | - | + | - | - | - |
| 11 | 12 | 13 | 14 | 15 | 16 | 17 | 18 | 19 | 20 |
| - | - | + | + | - | + | + | - | + | - |
| 21 | 22 | 23 | 24 | 25 | 26 | 27 | 28 | 29 | 30 |
| - | - | + | - | - | - | - | + | + | - |

*Note.* Developed by thesis's author.

Table 6 shows the complete data of the first five vehicles in each lane. We can see that the vehicles within the matched lanes have the same arriving and exiting times.

Table 6

*First Five Vehicles*

| Vehicle ID | Lane | Arrive Time | Right Turn | Exit Time |
|---|---|---|---|---|
| 123 | A1 | 0.0 | No | 17.179657557103365 |
| 206 | A2 | 0.0 | No | 17.179657557103365 |
| 301 | B1 | 1.0 | Yes | 18.179657557103365 |
| 413 | B2 | 1.0 | Yes | 18.179657557103365 |

*Table Continued*

| Vehicle ID | Lane | Arrive Time | Right Turn | Exit Time |
|---|---|---|---|---|
| 106 | A1 | 2.0 | No | 19.179657557103365 |
| 224 | A2 | 2.0 | No | 19.179657557103365 |
| 306 | B1 | 3.0 | Yes | 20.179657557103365 |
| 405 | B2 | 3.0 | Yes | 20.179657557103365 |
| 115 | A1 | 4.0 | Yes | 21.179657557103365 |
| 200 | A2 | 4.0 | Yes | 21.179657557103365 |
| 324 | B1 | 5.0 | No | 22.179657557103365 |
| 355 | B2 | 5.0 | No | 22.179657557103365 |
| 109 | A1 | 6.0 | Yes | 23.179657557103365 |
| 213 | A2 | 6.0 | Yes | 23.179657557103365 |
| 311 | B1 | 7.0 | No | 24.179657557103365 |
| 418 | B2 | 7.0 | No | 24.179657557103365 |
| 105 | A1 | 8.0 | No | 25.179657557103365 |
| 222 | A2 | 8.0 | No | 25.179657557103365 |
| 313 | B1 | 9.0 | Yes | 26.179657557103365 |
| 418 | B2 | 9.0 | Yes | 26.179657557103365 |

*Note.* Developed by thesis's author.

# CHAPTER 4

## CONCLUSION AND FUTURE WORK

In this thesis, we proposed a novel scheduling system that is based on the production line technique. In the proposed model, the intersection environment should be set up prior to the vehicles' arrival at the intersection. The goal of our system is to reduce the waiting time for the vehicles and to eliminate the chance of collision inside the intersection. Extensive simulations were carried out to evaluate three different traffic flow patterns (average, worst, and random). The K-Nearest Neighbor (KNN) method was used to predict the right turn vehicles. The results show that the proposed model provides high efficiency in the case of average and random pattern traffic flow. However, using extra space is still an issue in the worst-case traffic flow pattern; in this case the space always increases by 100%.

Our future work would consist of end-to-end scheduling of vehicles for the entire route. This is similar to the systems that airlines follow, where the route, speed, and time of departure and arrival are all set up and determined well in advance of taking off. For the next version of our model, we plan to incorporate the Internet of Things (IoT) paradigm including the sensors and data mining technology that can be used to determine the vehicles' arriving time precisely.

# REFERENCES

# APPENDICES

# APPENDIX A

# QUEUE FLOW CODING

```
package vflow2;

import java.util.ArrayList;
import java.util.*;
import java.util.Random;


public class VFlow2 {

    public static void main(String[] args) {

    System.out.println("The third pattern");


     int MaxQueueLong =720;
     int NumberofVehicles = 60; about
     double average=0.0;
     int runningTestingLoop =1;

     for(int test=0;test<runningTestingLoop;test++)
     {

     vehicle2 [] vehicle = new vehicle2[MaxQueueLong

     List <Integer> ArrangedArriving = new ArrayList();
     List <Integer> OriginalArriving = new ArrayList();
     List <Integer> TempArray = new ArrayList <Integer>(); //for temporary saving the
arriving times in order
     List <Integer> Road = new ArrayList();

     int newcount = 0;
     int ranking=0;
     int timesofLooping = 0;
     for(int loop=0;loop<=timesofLooping;loop++) //to insert the vehicle info
    {


    // 1 means there is  a request
    for(int i=0;i<MaxQueueLong;i++)  //this is the time
    {
             Road.add(randInt(0,1));
            // OriginalArriving.add(randInt(0,1));

    }
```

```
 System.out.println("The Original Road requests "+ Road);
 List <Integer> OrOr = new ArrayList();



 for(int ii=0;ii<Road.size();ii++)
{


   if(Road.get(ii)==1)
   {

        ArrangedArriving.add(ii);

        vehicle[newcount] = new vehicle2();
        vehicle[newcount].setArrTime(ii);
        vehicle[newcount].setRank(ranking);
        ranking++;
        newcount++;
   }    //ArrangedArriving.add(1);

}
}

System.out.println("----------------------------------------");

System.out.print("Explanation: The arrival time (the ranking)");
     System.out.print("\n");

     for(int jj=0;jj<newcount;jj++)
     {
         System.out.print( vehicle[jj].getArrTime()+ "("+ vehicle[jj].getRank()+")");

     }

// Now let's make the arranged array
List <Integer> ArrangedArray = new ArrayList <Integer>();
// 1 means there is  a request
int cc=0;
for(int i=0;i<newcount;i++){ //this is the time

    ArrangedArray.add(ArrangedArriving.get(i));  //take one arriving car
         ArrangedArray.add(0); // then add zero next to it
}
```

```
  System.out.println("\n\nThe Arranged Array "+ ArrangedArray);
  int xc=2; // to jump from one position to anothr in the arranged array
  int ss2=0;
   for(int qq=0;qq<newcount;qq++)
   {
         int temp1=vehicle[qq].getArrTime();
         int temp2 = ArrangedArray.get(ss2);
         int ind = ArrangedArray.indexOf(temp2);
         ss2+=2;
         vehicle[qq].setArrAfter(ind); //we sat the vehicles positions after arranging
them...we will  use that later to find the differences
   }



  // System.out.println("\n\nThe vehicles arriving times and after arranging time ");

   double countWaiting=0.0;
   double temp=0.0;


   for(int list=0;list<NumberofVehicles;list++)  // we took only the determined number of
vehicles
   {

       //to set waiting for each vehicle

       // now we chick the position of the vehicle in the orrignal array and in the arranged
one
       if(vehicle[list].getArrTime() > vehicle[list].getArrAfter())
       {
         // vehicle[list].setVehicleWaiting(vehicle[list].getArrTime() -
vehicle[list].getArrAfter());
          vehicle[list].setVehicleWaiting(0);
       }
       else if (vehicle[list].getArrTime() < vehicle[list].getArrAfter())
       {
          vehicle[list].setVehicleWaiting(vehicle[list].getArrAfter() -
vehicle[list].getArrTime());
       }
       else
         //if it didn't move and stayed at the same position in both arrays
       {
```

```
          vehicle[list].setVehicleWaiting(0);
        }
        // open the following one to see the calculation of the waiting time
        System.out.println("Vehicle number:" + list +".    Arrival and leaving time: " +
vehicle[list].getArrTime()+ "----" + vehicle[list].getArrAfter() + "  the waiting time:" +
vehicle[list].getVehicleWaiting());

       // countWaiting=(countWaiting*5.5880)+vehicle[list].getVehicleWaiting(); // we
collect the waiting time for every vehicle
          countWaiting=countWaiting+vehicle[list].getVehicleWaiting(); // we collect the
waiting time for every vehicle

    }
   //  System.out.println("\nThe count waiting " + countWaiting );
    // double AverageWaitingTime=(countWaiting*5.5880)/NumberofVehicles;

     temp = (countWaiting*5.5880); //here
     System.out.println("\nThe total waiting time for this loop is: " + temp + " seconds" );
     System.out.println("\nThe average waiting time for this loop is: " +
(temp/NumberofVehicles) + " seconds per vehicle" );

     average= average + (temp/NumberofVehicles); //add the aversge of this run to the
total average
     }

      System.out.println("\nThe average waiting time for WHOLE RUN is: " +
average/runningTestingLoop + " seconds per vehicle" );

 }
      public static int randInt(int min, int max)
  {

      Random rand = new Random();

      int randomNum = rand.nextInt((max - min) + 1) + min;

      return randomNum;
  }
}



/*
 * To change this license header, choose License Headers in Project Properties.
```

```
 * To change this template file, choose Tools | Templates
 * and open the template in the editor.
 */
package vflow2;

import static java.time.Clock.system;
import java.util.*;
/**
 *
 * @author User1
 */
public class vehicle2 {
    int a;
    int arivingTime;
    int VehicleRank;   //The rank (position) of the arrival vehicle in the queue
    int VehicleTimeAfter;
    int VehicletotalWaiting;

 public void setArrTime(int ar){   //We didn't use this...It's just an experiment
     this.arivingTime=ar;

 }

 public int getArrTime(){
    return this.arivingTime;
 }

 public void setRank(int rank){   // Ranking before sorting
     this.VehicleRank=rank;

 }

 public int getRank(){
    return this.VehicleRank;
 }

 public void setArrAfter(int timeAfter){   // Ranking before sorting
     this.VehicleTimeAfter=timeAfter;

 }

 public int getArrAfter(){
    return this.VehicleTimeAfter;
 }
```

```
 public void setVehicleWaiting(int vw){   //We didn't use this...It's just an experiment
    this.VehicletotalWaiting=vw;

 }

 public int getVehicleWaiting(){
   return this.VehicletotalWaiting;
 }


}
```

# APPENDIX B

# AIM TESTING CODE

```
package play1;

import java.util.ArrayList;
import java.util.Collections;
import java.util.List;
import java.util.Random;

public class tr4 {

	static double DistanceBP = 26.2467;  //Distance between two points in the graph 
feet
	static int numberofcars = 50; //maximum number 19*38

	//double totalFinal;
	static int numberofTestLoops = 1;
	static int numberofcarsX2 =numberofcars*2; //because of the diagonal, we 
calculated every collide twice

	//static double Total=ErrorsCount/TotalofMeetings*100;

	static double totalFinal;

	static double collectmyloops=0.0;

	//static double collectmyaccidents=0.0;
	static double CumErr=0.0;

	 public static void main(String[] args) {

		 double myloop=0.0;
		// double ErrorsCount=0.0;
		 double TotalofMeetings=0.0;
		 double TotalWaiting=0.0;
		double collectmyaccidents=0.0;
		for(int count=1;count<=numberofTestLoops;count++) {

			 double ErrorsCount=0.0;

			double CompleteHeight = 100.0; //foot
		double CompleteWidth = 100.0;	//foot

	       double The20percentofH = (CompleteHeight*(20.0/100.0));
	       double The20percentofW = (CompleteWidth*(20.0/100.0));
```

```
		System.out.println("The total height is : " + CompleteHeight );
		System.out.println("The total width is : " + CompleteWidth );

		System.out.println("20% of the width " + CompleteHeight  + " is " +
The20percentofH);
		System.out.println("20% of the height " + CompleteWidth  + " is " +
The20percentofW + "\n");



		   // Initialize cars
		Car3 [] Cars = new Car3[numberofcars];

		for(int CreatRandom=0;CreatRandom<Cars.length;CreatRandom++)
		{
		   // System.out.println("Cars. length" + Cars.length );

		Cars[CreatRandom] = new Car3();
		Cars[CreatRandom].setId(CreatRandom);

		Cars[CreatRandom].setSpeed(randInt(100,100));   //We sat their speed to
100 only
		}

	      // we prepare to put them in positions (x,y)
			//Make positions' arrays for vehicles going from west to east
			   List<Integer> listX1 = new ArrayList<>();
			for (int c1=1; c1<=38; c1++) { // to 38 because 39 is where the line
base, and 1 because 0 is the base line too
			listX1.add(c1);
			}


			List<Integer> listY1 = new ArrayList<>();
			for (int c2=40; c2<=58; c2++) {
			listY1.add(c2);
			}
			Collections.shuffle(listY1); // just to distribute them randomly in
different lanes


			List<Integer> listX2 = new ArrayList<>();
			for (int c2=40; c2<=58; c2++) {
			listX2.add(c2);
```

```
                }
                Collections.shuffle(listX2); // just to distribute them randomly in
different lanes

                List<Integer> listY2 = new ArrayList<>();
                for (int c1=1; c1<=38; c1++) { // to 38 because 39 is where the line
base, and 1 because 0 is the base line too
                listY2.add(c1);
                }



                int ir2=0;
                int ir4=0;

                int ir3=0;
                int ir5=0;
                int z=0;

          // we put them in directions and
         //We made it balance. Half of them goes through X axes and others through Y
axes
            for(int counter1=0;counter1<Cars.length/2;counter1++)
            {
                    Cars[counter1].setDirection(0); // going through the x axes
                        Cars[counter1].setTheX(listX1.get(ir2)); //set the X
                        Cars[counter1].setTheY(listY1.get(ir4)); //set the Y

                        ir2=ir2+1; //move to the next x cordinate

                         if (ir2==38)
                         {
                             ir2=0;  // this is to go back to the first
position in the suffle set
                            // ir2=ir2+1; //pick the next one from the
suffle set.
                             ir4=ir4+1; // move to the next y cordinate
                         }


            }
```

```
for(int counter2=Cars.length/2;
counter2<Cars.length;counter2++)
{
   Cars[counter2].setDirection(1); // going through the
y axes
   //Cars[counter0].setTheX(listX1.get(ir2)); //set the
X
   Cars[counter2].setTheX(listX2.get(ir3));
   Cars[counter2].setTheY(listY2.get(ir5));

   ir5=ir5+1;
   //z=z+1;
   //ir3=ir3+1;

   if(ir5==38)
    {

         //ir3=ir3+1; //here
         ir5=0;
         if(ir3<18
                    ){


         ir3=ir3+1;
         }


    }

}

// now lert's figure out the arriving times to the points that the vehicles
moving in the x axes will meet

for(int count3=0; count3<Cars.length;count3++ )
{
   int w = Cars[count3].getTheX();  //intial value = 15
   int li = Cars[count3].getTheY();  //          = 45
   System.out.print("\nThe meeting axes for car "
+(count3) + "("+ Cars[count3].getTheX() +","+ Cars[count3].getTheY() +")"+" are ");
   for (int v= 0;v<Cars.length;v++)
   {
         int searchXp=Cars[v].getTheX(); // The
second loop (when v=0) the value = 47
```

```
int searchYp=Cars[v].getTheY();  //
= 30
if
((w<searchXp&&searchXp>=40&&searchYp<=39))   //make it >39  // change li=39 to
{
System.out.print(" x:"+ searchXp
+"("+TimetoArrive(searchXp,w,Cars[count3].convFPS())+")");
Cars[count3].addPoint(searchXp); //
add the location axes to the vehicle's list
Cars[count3].addmCars(v); // add the
cars that will meet  /we don't need this anymore I guess
Cars[count3].addMymap(v,
TimetoArrive(searchXp,w,Cars[count3].convFPS())); // Cars[0] (1, 5.66690)

Cars[count3].addArriving(TimetoArrive(searchXp,w,Cars[count3].convFPS()));
//add the point reaching time
//TotalofMeetings++;
}

if
((li<searchYp&&searchYp>=40&&w>=39))  // make it 40
{
System.out.print(" y:"+ searchYp
+"("+TimetoArrive(searchYp,li,Cars[count3].convFPS())+")");
Cars[count3].addPoint(searchYp);
Cars[count3].addmCars(v); // add the
cars that will meet
Cars[count3].addMymap(v,
TimetoArrive(searchYp,li,Cars[count3].convFPS()));

Cars[count3].addArriving(TimetoArrive(searchYp,li,Cars[count3].convFPS()));
//TotalofMeetings++;
}

}
```

```
                for(int list=0;list<Cars.length;list++)
                {
                   //+ " Y:" + Cars[list].getTheY()
                // System.out.println("vehicle number " + Cars[list].getId() +" Direction "
+Cars[list].getDirection() + " X:" + Cars[list].getTheX()+ " Y:" + Cars[list].getTheY());

                }

                                    }





        /***********************************************************/


System.out.println("\n\nThe indexes cars that car 1 will meet : " + Cars[0].getmCars());
                                    System.out.println("Meeting points for car 1 : " +
Cars[0].getPoints());
                                    System.out.println("Arriving times for car 1 to these points
: " + Cars[0].getArriving());
                                    System.out.println("How long the car will stay in any point
: " + Cars[0].getPoOccTime());


                                    for (int op0=0;op0<numberofcars;op0++)
                                    {
                                            //we got all the cars that the current car will meet

                                      for(int op1=0;op1<Cars.length;op1++)
                                      {


if(Cars[op0].getDirection()==0&&Cars[op1].getDirection()==1)
                                            {

                                      System.out.println("\n Car"+(op0)+ "
["+Cars[op0].getTheX()+","+ Cars[op0].getTheY()+"]");
                                          double CarR =
TimetoArrive(Cars[op1].getTheX(),Cars[op0].getTheX(),Cars[op0].convFPS());
```

```
                                        double LeavingPT = CarR+Cars[op0].getPoOccTime();
//the leaving time for car1
                                        double CarR2 =
TimetoArrive(Cars[op0].getTheY(),Cars[op1].getTheY(),Cars[op1].convFPS());
                                        double LeavingPT2 =
CarR2+Cars[op1].getPoOccTime(); ////the leaving time for car2


                                        double TempVariable1 = Cars[op0].getMymap(op1);
//the arriving time for car1
                                        double TempVariable2 = Cars[op1].getMymap(op0);
//the arriving time for car2

                                           System.out.print(" compare "+TempVariable1 +" - "+
LeavingPT +" with " +TempVariable2+" - "+ LeavingPT2 );

if(TempVariable1>LeavingPT2||LeavingPT<TempVariable2) //if any car left before the
arrival of the other car, then there is no error
                                       {

                                             System.out.println(" = No Error");
                                             TotalofMeetings++;
                                             System.out.println("\n");
                                       }
                                       else  {
                                             System.out.println(" = There is an Error");
                                             ErrorsCount++;
                                             TotalofMeetings++;
                                             TotalWaiting= TotalWaiting+(146.667/26.2467);
//speed is 100MPH
                                             System.out.print(" Vehicles locations: ("
+Cars[op0].getTheX()+ ","+Cars[op0].getTheY()+")" + " and (" + Cars[op1].getTheX() +
"," +Cars[op1].getTheY()+")");

                                                  double GetTheVehicleCurrentWaitingTime
= Cars[op0].getWaiting();
                                                  //int AddToTheVehicleWaitingtime=

Cars[op0].setWaiting(GetTheVehicleCurrentWaitingTime+5.58);
                                                   System.out.print(" my waaaaaating after
this meeting"+ Cars[op0].getWaiting());
                                                   System.out.println("\n");
```

```
for(int iii=0;iii<Cars.length;iii++) //add
waiting times to the vehicles behind
{
if
(Cars[iii].getTheX()<=Cars[op0].getTheX()&&Cars[iii].getTheY()==Cars[op0].getTheY
()) //any vehicle with the same Y but less X will get a waiting time
{
double
GetTheLinesVehicleCurrentWaitingTime = Cars[iii].getWaiting();

Cars[iii].setWaiting(GetTheLinesVehicleCurrentWaitingTime+5.58);
}

}


}
//TotalofMeetings++;


}


if(Cars[op0].getDirection()==1&&Cars[op1].getDirection()==0)
{

System.out.println("\n Car"+(op0));
System.out.println("\n Car"+(op0)+ "
["+Cars[op0].getTheX()+","+ Cars[op0].getTheY()+"]");
double CarR =
TimetoArrive(Cars[op1].getTheY(),Cars[op0].getTheY(),Cars[op0].convFPS());
double LeavingPT = CarR+Cars[op0].getPoOccTime();
//the leaving time for car1
double CarR2 =
TimetoArrive(Cars[op0].getTheX(),Cars[op1].getTheX(),Cars[op1].convFPS());
double LeavingPT2 =
CarR2+Cars[op1].getPoOccTime(); ////the leaving time for car2



double TempVariable1 = Cars[op0].getMymap(op1);
//the arriving time for car1
double TempVariable2 = Cars[op1].getMymap(op0);
//the arriving time for car2
```

```
                              //Compare the arriving and leaving times for the two cars
                                System.out.print(" compare "+TempVariable1 +" - "+
LeavingPT +" with " +TempVariable2+" - "+ LeavingPT2 );

if(TempVariable1>LeavingPT2||LeavingPT<TempVariable2)
                              {
                                    System.out.println(" = No Error");
                                    TotalofMeetings++;
                                    System.out.println("\n");
                              }
                              else {
                                    System.out.println(" = There is an Error");
                                    ErrorsCount++;
                                    TotalofMeetings++;
                                    TotalWaiting= TotalWaiting+(146.667/26.2467);
                                    System.out.print(" Vehicles locations: ("
+Cars[op0].getTheX()+ ","+Cars[op0].getTheY()+")" + " and (" + Cars[op1].getTheX() +
"," +Cars[op1].getTheY()+")");

                                    double GetTheVehicleCurrentWaitingTime2 =
Cars[op0].getWaiting();

Cars[op0].setWaiting(GetTheVehicleCurrentWaitingTime2+5.58);
                                          System.out.print(" my waaaaaating after
this meeting"+ Cars[op0].getWaiting());
                                     System.out.println("\n");

                                     for(int iii2=0;iii2<Cars.length;iii2++) //add
waiting times to the vehicles behind
                                          {
                                                if
(Cars[iii2].getTheX()==Cars[op0].getTheX()&&Cars[iii2].getTheY()<=Cars[op0].getTh
eY()) //any vehicle with the same Y but less X will get a waiting time
                                                {
                                                      double
GetTheLinesVehicleCurrentWaitingTime2 = Cars[iii2].getWaiting();

Cars[iii2].setWaiting(GetTheLinesVehicleCurrentWaitingTime2+5.58);
                                                }

                                          }

                                 }
```

```
                    }

                }

            }

            myloop=((double)ErrorsCount/2.00)/(double)
numberofcars *100.00; // divided by 2 because I count them twice before

System.out.println("*****************************************************
");
           // System.out.println("This "+ count +" Loop Result:
"+myloop+ "% Error Rate" );
            System.out.println("\n");
            System.out.println("AVERAGE EXPECTED
ACCIDENTS  "+((ErrorsCount/2.00)/numberofcars) + " for " + numberofcars + " cars" );

            //but we only took the average
            collectmyaccidents = collectmyaccidents
+((ErrorsCount/2.00)/numberofcars);

            //  System.out.println("\n");
          //  System.out.println("DIVIDED BY NUMBER OF CARS
"+(ErrorsCount/2.00)/numberofcars);

            System.out.println("\n");

           // CumErr=CumErr+myloop;




            int finalcountwaiting;
                double collect=0.0;
                for(int cw=0;cw<Cars.length;cw++)
                {
                    collect = collect+Cars[cw].getWaiting();
                }
```

```
                                        System.out.println("\n THE COLLECTED
WAITING TIMES IS " + collect/2.0 + " SECONDS"); //because they are two sides...only
one of the wait
                                        System.out.println("\n");
                                        System.out.println("\n THE AVERAGE
WAITING TIMES FOR THIS LOOP IS " + (collect/2.0)/numberofcars + " SECONDS
PER VEHICLE"); // all of them now because the study is to take into account the total
number of cars in the intersection
                                        System.out.println("\n");
                                        collectmyloops =
collectmyloops+((collect/2.0)/numberofcars);

                                        //System.out.println("AVERAGE EXPECTED
ACCIDENTS  "+((ErrorsCount/2.00)/numberofcars) + " for " + numberofcars + " cars" );

                                       //but we only took the average
                                      //  collectmyaccidents = collectmyaccidents
+((ErrorsCount/2.00)/numberofcars);



                } //finish the loop


        System.out.println("************************************************
****");

        System.out.println("************************************************
****");

        System.out.println("************************************************
****");

        System.out.println("************************************************
****");

        System.out.println("************************************************
****");

                System.out.println("The Final Average waiting for the WHOLE RUN : "
+collectmyloops/(double)numberofTestLoops + " SECONDS PER VEHICLE");
```

```
                System.out.println("The Final Average expected collisions for the
WHOLE RUN : " +collectmyaccidents/(double)numberofTestLoops + " PER
VEHICLE");

        }



        public static int randInt(int min, int max)
          {

              Random rand = new Random();

              int randomNum = rand.nextInt((max - min) + 1) + min;

              return randomNum;
          }

        public static double TimetoArrive(int a, int b, int c)
          {
          // a and b are x and y
                // c is the speed
              double Seconds = (((a-b)*DistanceBP)/c) ; //how long does it take a car to
arrive to the meeting point;

              return Seconds;
          }

}




package play1;

import java.util.ArrayList;
import java.util.HashMap;
import java.util.Map;

public class Car3 {

        private int speed;  // the speed of the car
        private int feetPerSecond; // convert from mile/hour to feet/second
```

```
        private int TheX;
        private int TheY;
        private int id;
        private double CarWaiting;
        private int Direction; //0 going to the X axes , 1 going to the Y axes
        private int TheMeetingPoints;
        private double PoOccTime; //the time the car needs to occupy a point
        private ArrayList<Integer> mpoints = new ArrayList<Integer>(); //the meeting
points
        private ArrayList<Integer> mcars = new ArrayList<Integer>(); //the meeting cars
        private ArrayList<Double> TheArrivingTime = new ArrayList<Double>(); //at
what time the car will arrive
        private HashMap<Integer, Double> myMap = new HashMap <Integer,
Double>();
        //private ArrayList<Integer> mcars2 = new ArrayList<Integer>(); //the meeting
cars
        //myMap.put(courseID, scores);
        // 1 , 2 , 5.7952
        public int setSpeed(int s) {

                this.speed=s;
                 return s;
        }

        public int getSpeed() {
                return this.speed;
   }

  public int setId(int id) {

                this.id=id;
                 return id;
        }

        public int getId() {
                return this.id;
   }

public double setWaiting(double wait) {

                this.CarWaiting=wait;
                 return wait;
        }
```

```
	public double getWaiting() {
		return this.CarWaiting;
}
	public int convFPS(){

		this.feetPerSecond = (((speed*5280)/60)/60);
		return feetPerSecond; //convert from mile to feet


	}
	public void setTheX(int x){
		this.TheX=x;
	}
	public int getTheX(){
		return this.TheX;

	}
	public void setTheY(int y){
		this.TheY=y;
	}
	public int getTheY(){
		return this.TheY;
	}

	public void setDirection(int d){
		this.Direction=d;
	}
	public int getDirection(){
		return this.Direction;
	}

	public void addmCars(int a){
		this.mcars.add(a);
	}

	public ArrayList<Integer> getmCars(){
		return this.mcars;
	}
	public void addPoint(int a1){
		this.mpoints.add(a1);
	}
	public ArrayList<Integer> getPoints(){
		return this.mpoints;
	}
```

```
    public void addArriving(Double a2){
        this.TheArrivingTime.add(a2);
    }
    public ArrayList<Double> getArriving(){
        return this.TheArrivingTime;
    }
    public void addMymap(int a3, double b3){
         this.myMap.put(a3,b3);
    }
    public  Double getMymap(int w){
        return this.myMap.get(w);
    }
    public double getPoOccTime(){
        return this.PoOccTime =26.2467/this.convFPS();
    }

}
```

# APPENDIX C

# PRODUCTION LINE SCHEDULING CODE

```
package play2;

        import java.io.BufferedWriter;
        import java.io.File;
        import java.io.FileWriter;
        import java.io.IOException;
        import java.io.PrintWriter;
        import java.util.*;


        import java.io.BufferedReader;
        import java.io.BufferedWriter;
        import java.io.File;
        import java.io.FileNotFoundException;
        import java.io.FileReader;
        import java.io.FileWriter;
        import java.io.IOException;
        import java.io.InputStream;
        import java.io.InputStreamReader;
        import java.io.PrintWriter;
        import java.nio.charset.StandardCharsets;
        import java.nio.file.Files;
        import java.nio.file.Paths;
        import java.util.*;
```

```
import javax.swing.JOptionPane;


//import play2.KNN.Turn;

import play2.KNN.DistanceComparator;

import play2.KNN.Result;

import static jdk.nashorn.tools.ShellFunctions.input;

//import play1.Car;

// The K nearest neighbor code has been modified to match the needs:

// imported from DR NOUREDDIN SADAWI

 //http://www.imperial.ac.uk/people/n.sadawi

// https://github.com/nsadawi/KNN

// https://www.youtube.com/channel/UCNYv4HA3WjV3gZGLfBehRWQ


public class tr2

{


        static int numberofSpots=60;

        static double MinLaneA1Speed=60.00;

        static double MaxLaneA1Speed=65.00;

        static double MinLaneA2Speed=60.00;

        static double MaxLaneA2Speed=65.00;

        static double MinLaneB1Speed=60.00;
```

```
static double MaxLaneB1peed=65.00;

static double MinLaneB2Speed=60.00;

static double MaxLaneB2peed=65.00;

static int LaneB1Speed=10;

static double SpotsLength=26.2467; // 8 meter

static int SpotOccupation;

static int RightTurnForA=2;

static int RightTurnForB=3;  // 100 means there is nothing


double ErrorsCount=0.0;

double LoopResult=0.0;


int CompleteHeight = 100;

int CompleteWidth = 100;


int The20percentofH = (int)(CompleteHeight*(20.0f/100.0f));

int The20percentofW = (int)(CompleteWidth*(20.0f/100.0f));




public static void main(String[] args) throws

InterruptedException,FileNotFoundException, IOException
```

```
{

            Car2 [] A1 = new Car2[numberofSpots/2];  //Lane A1    // % by 2
because only half of the spots will be occupied

            Car2 [] A2 = new Car2[numberofSpots/2];  //Lane A2

            Car2 [] B1 = new Car2[numberofSpots/2]; //Lane B1

            Car2 [] B2 = new Car2[numberofSpots/2]; //Lane B1

           //Lane A1 saving arrays
            List<Double> ArrivaltimesArray = new ArrayList<>();
            List<Double> StayingtimesArray = new ArrayList<>();
            List<Double> ExitingtimesArray = new ArrayList<>();
            List<Double> ExitingtimesArrayTurn = new ArrayList<>();
            List<Integer> RightTurnArray = new ArrayList<>();
            List<String> RightTTurnArray = new ArrayList<>();
           // List<Integer> A1 = new ArrayList<>();
```

```
//Lane A2 saving arrays
  List<Double> ArrivaltimesArrayA2 = new ArrayList<>();
  List<Double> StayingtimesArrayA2 = new ArrayList<>();
  List<Double> ExitingtimesArrayA2 = new ArrayList<>();
  List<Double> ExitingtimesArrayTurnA2 = new ArrayList<>();
  List<Integer> RightTurnArrayA2 = new ArrayList<>();
  List<String> RightTTurnArrayA2 = new ArrayList<>();

//Lane B1 saving arrays
  List<Double> ArrivaltimesArray2 = new ArrayList<>();
  List<Double> StayingtimesArray2 = new ArrayList<>();
  List<Double> ExitingtimesArray2 = new ArrayList<>();
  List<Double> ExitingtimesArrayTurn2 = new ArrayList<>();
  List<Integer> RightTurnArray2 = new ArrayList<>();
  List<String> RightTTurnArray2 = new ArrayList<>();

//Lane B2 saving arrays
  List<Double> ArrivaltimesArrayB2 = new ArrayList<>();
  List<Double> StayingtimesArrayB2 = new ArrayList<>();
  List<Double> ExitingtimesArrayB2 = new ArrayList<>();
  List<Double> ExitingtimesArrayTurnB2 = new ArrayList<>();
```

```
List<Integer> RightTurnArrayB2 = new ArrayList<>();
List<String> RightTTurnArrayB2 = new ArrayList<>();

//A1 seconds when its open
  List<Integer> listSeconds = new ArrayList<>();
  for (int ir=0; ir<=59; ir+=2) {
          listSeconds.add(ir);
  }
 // Collections.shuffle(listSeconds); //to enter 1 car per 1 second, if we
didn't use the shuffle at all andinstead we generate a random number again we might take
the same number again

  //A1 vehicles ID
  List<Integer> A1Names = new ArrayList<>();
  for (int A1N=100; A1N<=129; A1N++) {
          A1Names.add(A1N);
  }
  Collections.shuffle(A1Names);


//A2 vehicles Names
 List<Integer> A2Names = new ArrayList<>();
```

```
  for (int A2N=200; A2N<=229; A2N++) {
        A2Names.add(A2N);
  }
  Collections.shuffle(A2Names);


//A2
  List<Integer> listSecondsA2 = new ArrayList<>();
  for (int irA2=0; irA2<=59; irA2+=2) {
        listSecondsA2.add(irA2);
  }


  //B1
  List<Integer> listSeconds2 = new ArrayList<>();
  for (int ir2=1; ir2<=59; ir2+=2) {
        listSeconds2.add(ir2);
  }


  //B1 names
  List<Integer> B1Names = new ArrayList<>();
```

```
for (int B1N=300; B1N<=329; B1N++) {
        B1Names.add(B1N);
}
Collections.shuffle(B1Names);




//B2
List<Integer> listSecondsB2 = new ArrayList<>();
for (int irB2=1; irB2<=59; irB2+=2) {
        listSecondsB2.add(irB2);
}

//Collections.shuffle(listSecondsB2); //to enter 1 car per 1 second, if we
didn't use the shuffle they might take the same value again

//B1 names
List<Integer> B2Names = new ArrayList<>();
for (int B2N=400; B2N<=429; B2N++) {
        B2Names.add(B2N);
}
Collections.shuffle(B2Names);
```

```
//A1
for(int CreatRandom=0;CreatRandom<A1.length;CreatRandom++)
{
    // if rightturnforA ==1; then go here

    {

        A1[CreatRandom] = new Car2();
        A1[CreatRandom].setSpeed(randInt(60,65));
        A1[CreatRandom].setDirection(01);
        A1[CreatRandom].setEntering(0);
        A1[CreatRandom].setTurn(RightTurnForA);
        A1[CreatRandom].setVID(A1Names.get(CreatRandom));

A1[CreatRandom].setGateArrivalTime(listSeconds.get(CreatRandom));
```

```
A1[CreatRandom].setStayingTime((26.2467*numberofSpots)/91.66667);  //time =
distance/speed    the 91.66667 is the converting of 62.5 MPH to foot per second

A1[CreatRandom].setExitTime(A1[CreatRandom].getGateArrivalTime()+
A1[CreatRandom].getStayingTime());

ArrivaltimesArray.add(A1[CreatRandom].getGateArrivalTime());

StayingtimesArray.add(A1[CreatRandom].getStayingTime());
                    ExitingtimesArray.add(A1[CreatRandom].getExitTime());

ExitingtimesArrayTurn.add(A1[CreatRandom].getExitTime());
                    RightTurnArray.add(A1[CreatRandom].getTurn());

                    A1[CreatRandom].setDay(randInt(1,5));
                    A1[CreatRandom].setHour(randInt(0,23));
                    A1[CreatRandom].setEvent(randInt(0,1));

                // Read the entire file in
```

```
List<String> myFileLines =
Files.readAllLines(Paths.get("C:\\Users\\User1\\Desktop\\Research\\The
Paper\\instnces.txt"));

// Remove any blank lines
for (int i = myFileLines.size() - 1; i >= 0; i--) {
    if (myFileLines.get(i).isEmpty()) {
        myFileLines.remove(i);
    }
}

// Declare you 2d array with the amount of lines that were read
from the file
int[][] intArray = new int[myFileLines.size()][];

// Iterate through each row to determine the number of columns
for (int i = 0; i < myFileLines.size(); i++) {
    // Split the line by spaces
    String[] splitLine = myFileLines.get(i).split("\\s");

    // Declare the number of columns in the row from the split
    intArray[i] = new int[splitLine.length];
```

```
    for (int j = 0; j < splitLine.length; j++) {
        // Convert each String element to an integer
        intArray[i][j] = Integer.parseInt(splitLine[j]);
       // dataarray[i][j] = Integer.parseInt(splitLine[j]);
     // dataarray
    }
}

// Print the integer array
for (int[] row : intArray) {
    for (int col : row) {
        System.out.printf("%5d ", col);
    }
    System.out.println();
}

       // Read the entire file in

// Read the the turning file file in
List<String> myFileLinesTurn =
Files.readAllLines(Paths.get("C:\\Users\\User1\\Desktop\\Research\\The
Paper\\Turn.txt"));
```

```
    File file = new File("C:\\Users\\User1\\Desktop\\Research\\The
Paper\\Turn.txt");
Scanner input = new Scanner(file);
List<String> list0 = new ArrayList<String>();

while (input.hasNextLine()) {
  list0.add(input.nextLine());
}



System.out.println("The classification of the new vehicle: "+"\n");
//System.out.println(intArray[1][1]);
                int k = 3;// # of neighbours
                //list to save turn data
                List<Turn> turnList = new ArrayList<Turn>();
                //list to save distance result
                List<Result> resultList = new ArrayList<Result>();
                // add turn data to turnList

            int dayy =  A1[CreatRandom].getDay();
      int hourr =  A1[CreatRandom].getHour();
```

```
int eventt=  A1[CreatRandom].getEvent();




int intialloop = myFileLines.size() - 1; //add to this whenever
a new element added
// int intialloop =  CreatRandom
int innerintialloop = 0;
  int[] query = {dayy,hourr,eventt};
  clearfileInstance();
  for (int loop1=0;loop1<=intialloop;loop1++)
  {
      for(int loop2=0;loop2<=innerintialloop;loop2++)
      {

          WriteInstance(intArray [loop1][loop2],intArray
[loop1][loop2+1],intArray [loop1][loop2+2]);

      }

  }
     WriteInstance(query[0],query[1],query[2]);
```

```
//because different types

for(int aabb = 0; aabb <list0.size(); aabb++)
   {
  // System.out.println(list0.get(aabb));
   turnList.add(new
Turn(intArray[aabb],(list0.get(aabb))));

   }

        //find disnaces
        for(Turn turn : turnList){
                double dist = 0.0;
                for(int j = 0; j < turn.turnAttributes.length;
j++){
                        dist +=
Math.pow(turn.turnAttributes[j] - query[j], 2) ;

        //System.out.print(turn.turnAttributes[j]+" ");
                }
                double distance = Math.sqrt( dist );
```

```
                    resultList.add(new
Result(distance,turn.turnSign));
                    //System.out.println(distance);
                }

                //System.out.println(resultList);
                Collections.sort(resultList, new
DistanceComparator());
                String[] ss = new String[k];
                for(int x = 0; x < k; x++){

    System.out.println(resultList.get(x).distance+" ("+ resultList.get(x).turnSign+
")");
                    //get classes of k nearest instances (turn
names) from the list into an array
                    ss[x] = resultList.get(x).turnSign;
                }
                String majClass = findMajorityClass(ss);

                System.out.println("The class of the new vehicle is :
("+majClass+")");
            //System.out.println("The [K] is  : "+k+"");
```

```
WriteTurn(majClass);

A1[CreatRandom].setTTurn(majClass);

RightTTurnArray.add( A1[CreatRandom].getTTurn());


// WriteK(k);

}




}




//A2

for(int

CreatRandomA2=0;CreatRandomA2<A2.length;CreatRandomA2++)

{


{


A2[CreatRandomA2] = new Car2();

A2[CreatRandomA2].setSpeed(randInt(60,65));

A2[CreatRandomA2].setDirection(02);

A2[CreatRandomA2].setEntering(0);
```

```
A2[CreatRandomA2].setTurn(RightTurnForA);

A2[CreatRandomA2].setVID(A2Names.get(CreatRandomA2));

A2[CreatRandomA2].setGateArrivalTime(listSecondsA2.get(CreatRandomA2));

A2[CreatRandomA2].setStayingTime((26.2467*numberofSpots)/91.66667);  //time = distance/speed    the 91.66667 is the converting of 62.5 MPH to foot per second

A2[CreatRandomA2].setExitTime(A2[CreatRandomA2].getGateArrivalTime()+ A2[CreatRandomA2].getStayingTime());

ArrivaltimesArrayA2.add(A2[CreatRandomA2].getGateArrivalTime());

StayingtimesArrayA2.add(A2[CreatRandomA2].getStayingTime());

ExitingtimesArrayA2.add(A2[CreatRandomA2].getExitTime());

ExitingtimesArrayTurnA2.add(A2[CreatRandomA2].getExitTime());
RightTurnArrayA2.add(A2[CreatRandomA2].getTurn());
```

```
A2[CreatRandomA2].setDay(A1[CreatRandomA2].getDay());

A2[CreatRandomA2].setHour(A1[CreatRandomA2].getHour());

A2[CreatRandomA2].setEvent(A1[CreatRandomA2].getEvent());

A2[CreatRandomA2].setTTurn(A1[CreatRandomA2].getTTurn());
                    RightTTurnArrayA2.add(A2[CreatRandomA2].getTTurn());
                      }

                }

               //B1
               for(int CreatRandom2=0;CreatRandom2<B1.length;CreatRandom2++)
                {
                     //we could use the % but we have an issue with the 0
                      {

                            B1[CreatRandom2] = new Car2();
                        B1[CreatRandom2].setSpeed(randInt(60,65));
```

```
B1[CreatRandom2].setDirection(11);
B1[CreatRandom2].setEntering(0);
B1[CreatRandom2].setTurn(RightTurnForB);

B1[CreatRandom2].setVID(B1Names.get(CreatRandom2));

B1[CreatRandom2].setGateArrivalTime(listSeconds2.get(CreatRandom2));

B1[CreatRandom2].setStayingTime((26.2467*numberofSpots)/91.66667);  //time =
distance/speed    the 91.66667 is the converting of 62.5 MPH to foot per second

B1[CreatRandom2].setExitTime(B1[CreatRandom2].getGateArrivalTime()+
B1[CreatRandom2].getStayingTime());

ArrivaltimesArray2.add(B1[CreatRandom2].getGateArrivalTime());

StayingtimesArray2.add(B1[CreatRandom2].getStayingTime());

ExitingtimesArray2.add(B1[CreatRandom2].getExitTime());

ExitingtimesArrayTurnA2.add(B1[CreatRandom2].getExitTime());
RightTurnArray2.add(B1[CreatRandom2].getTurn());
```

```
B1[CreatRandom2].setDay(randInt(1,5));
B1[CreatRandom2].setHour(randInt(0,23));
B1[CreatRandom2].setEvent(randInt(0,1));

 // Read the entire file in
List<String> myFileLinesB =
Files.readAllLines(Paths.get("C:\\Users\\User1\\Desktop\\Research\\The
Paper\\instncesB.txt"));

// Remove any blank lines
for (int i = myFileLinesB.size() - 1; i >= 0; i--) {
    if (myFileLinesB.get(i).isEmpty()) {
            myFileLinesB.remove(i);
    }
}

// Declare you 2d array with the amount of lines that were
read from the file
int[][] intArrayB = new int[myFileLinesB.size()][];
```

```
        // Iterate through each row to determine the number of
columns
        for (int i = 0; i < myFileLinesB.size(); i++) {
            // Split the line by spaces
            String[] splitLine = myFileLinesB.get(i).split("\\s");

            // Declare the number of columns in the row from the
split
            intArrayB[i] = new int[splitLine.length];
            for (int j = 0; j < splitLine.length; j++) {
                // Convert each String element to an integer
                    intArrayB[i][j] = Integer.parseInt(splitLine[j]);
              // dataarray[i][j] = Integer.parseInt(splitLine[j]);
             // dataarray
            }
        }

        // Print the integer array
        for (int[] row : intArrayB) {
            for (int col : row) {
                System.out.printf("%5d ", col);
            }
```

```
System.out.println();
}

// Read the entire file in

// Read the the turning file file in
List<String> myFileLinesTurn =
Files.readAllLines(Paths.get("C:\\Users\\User1\\Desktop\\Research\\The
Paper\\TurnB.txt"));

File file = new
File("C:\\Users\\User1\\Desktop\\Research\\The Paper\\TurnB.txt");
Scanner input = new Scanner(file);
List<String> list0 = new ArrayList<String>();

while (input.hasNextLine()) {
  list0.add(input.nextLine());
}

System.out.println("The classification of the new vehicle:
"+"\n");
//System.out.println(intArray[1][1]);
```

```
int k = 3;// # of neighbours
//list to save turn data
List<Turn> turnList = new
ArrayList<Turn>();
//list to save distance result
List<Result> resultList = new
ArrayList<Result>();
// add Turn data to turnList



int dayy =  B1[CreatRandom2].getDay();
int hourr =  B1[CreatRandom2].getHour();
int eventt=  B1[CreatRandom2].getEvent();




int intialloop = myFileLinesB.size() - 1; //add to this
whenever a new element added
// int intialloop =  CreatRandom
int innerintialloop = 0;
int[] query = {dayy,hourr,eventt};
clearfileInstanceB();
```

```
for (int loop1=0;loop1<=intialloop;loop1++)
{
    for(int loop2=0;loop2<=innerintialloop;loop2++)
    {

        WriteInstanceB(intArrayB
[loop1][loop2],intArrayB [loop1][loop2+1],intArrayB [loop1][loop2+2]);

    }

}
  WriteInstanceB(query[0],query[1],query[2]);
 //because different types

 for(int aabb = 0; aabb <list0.size(); aabb++)
     {
    // System.out.println(list0.get(aabb));
     turnList.add(new
Turn(intArrayB[aabb],(list0.get(aabb))));

     }
```

```
//find disnaces
for(Turn turn : turnList){
    double dist = 0.0;
    for(int j = 0; j <
turn.turnAttributes.length; j++){
        dist +=
Math.pow(turn.turnAttributes[j] - query[j], 2) ;

    //System.out.print(turn.turnAttributes[j]+" ");
    }
    double distance = Math.sqrt( dist );
    resultList.add(new
Result(distance,turn.turnSign));
    //System.out.println(distance);
}

//System.out.println(resultList);
Collections.sort(resultList, new
DistanceComparator());
String[] ss = new String[k];
for(int x = 0; x < k; x++){
```

```
System.out.println(resultList.get(x).distance+" ("+ resultList.get(x).turnSign+
")");
                                        //get classes of k nearest instances
(turn names) from the list into an array
                                        ss[x] = resultList.get(x).turnSign;
                                    }
                                    String majClass = findMajorityClass(ss);

                                    System.out.println("The class of the new
vehicle is : ("+majClass+")");
                                //System.out.println("The [K] is  : "+k+"");
                                      WriteTurnB(majClass);
                                 B1[CreatRandom2].setTTurn(majClass);
                                RightTTurnArray2.add(
B1[CreatRandom2].getTTurn());

                                // WriteK(k);

                    }
```

```
}

for(int
CreatRandomB2=0;CreatRandomB2<B2.length;CreatRandomB2++)
{

{

B2[CreatRandomB2] = new Car2();
B2[CreatRandomB2].setSpeed(randInt(60,65));
B2[CreatRandomB2].setDirection(12);
B2[CreatRandomB2].setEntering(0);
B2[CreatRandomB2].setTurn(RightTurnForB);

B2[CreatRandomB2].setVID(B2Names.get(CreatRandomB2));

B2[CreatRandomB2].setGateArrivalTime(listSecondsB2.get(CreatRandomB2));

B2[CreatRandomB2].setStayingTime((26.2467*numberofSpots)/91.66667);  //time =
distance/speed    the 91.66667 is the converting of 62.5 MPH to foot per second
```

```
B2[CreatRandomB2].setExitTime(B2[CreatRandomB2].getGateArrivalTime()+
B2[CreatRandomB2].getStayingTime());

ArrivaltimesArrayB2.add(B2[CreatRandomB2].getGateArrivalTime());

StayingtimesArrayB2.add(B2[CreatRandomB2].getStayingTime());

ExitingtimesArrayB2.add(B2[CreatRandomB2].getExitTime());

ExitingtimesArrayTurnB2.add(B2[CreatRandomB2].getExitTime());
                    RightTurnArrayB2.add(B2[CreatRandomB2].getTurn());



B2[CreatRandomB2].setDay(B1[CreatRandomB2].getDay());

B2[CreatRandomB2].setHour(B1[CreatRandomB2].getHour());

B2[CreatRandomB2].setEvent(B1[CreatRandomB2].getEvent());

B2[CreatRandomB2].setTTurn(B1[CreatRandomB2].getTTurn());
```

```
RightTTurnArrayB2.add(B2[CreatRandomB2].getTTurn());

    }

}




    System.out.println("The arrival times for the vehicles on A1 axes
are: "+ ArrivaltimesArray + " ");
    System.out.println("The staying times for the vehicles on A1 axes
are: "+ StayingtimesArray + " ");
    System.out.println("The Exiting times for the vehicles on A1 axes
are: "+ ExitingtimesArray + " ");
    System.out.println("Right turn arry for the vehicles on A1 axes
are: "+ RightTTurnArray + " ");
    System.out.println("A1 IDs are are: "+ A1Names + " ");
    System.out.println("");
   // System.out.println();

    System.out.println("The arrival times for the vehicles on A2 axes
are: "+ ArrivaltimesArrayA2 + " ");
```

```
System.out.println("The staying times for the vehicles on A2 axes
are: "+ StayingtimesArrayA2 + " ");
System.out.println("The Exiting times for the vehicles on A2 axes
are: "+ ExitingtimesArrayA2 + " ");
System.out.println("The right turn arry for the vehicles on A2
axes are: "+ RightTTurnArrayA2 + " ");
System.out.println("A2 IDs are are: "+ A2Names + " ");
System.out.println("");

System.out.println("The arrival times for the vehicles on B1 axes
are: "+ ArrivaltimesArray2 + " ");
System.out.println("The staying times for the vehicles on B1 axes
are: "+ StayingtimesArray2 + " ");
System.out.println("The Exiting times for the vehicles on B1 axes
are: "+ ExitingtimesArray2 + " ");
System.out.println("The right turn arry for the vehicles on B1
axes are: "+ RightTTurnArray2 + " ");
System.out.println("B1 IDs are are: "+ B1Names + " ");
System.out.println("");

System.out.println("The arrival times for the vehicles on B2 axes
are: "+ ArrivaltimesArrayB2 + " ");
```

```
                        System.out.println("The staying times for the vehicles on B2 axes
are: "+ StayingtimesArrayB2 + " ");
                        System.out.println("The Exiting times for the vehicles on B2 axes
are: "+ ExitingtimesArrayB2 + " ");
                        System.out.println("The right turn arry for the vehicles on B2
axes are: "+ RightTTurnArrayB2 + " ");
                        System.out.println("B2 IDs are are: "+ B2Names + " ");
                        System.out.println("");




          int timet= 1 * 0; // Convert to seconds
          long delay = timet * 1000; //5000 milliseconds equal to(5 seconds).



          do
          {
           // int minutes = timet / 60;
            int seconds = timet % 60;
            System.out.println( seconds + " second(s)");

           // IF IT MEETS THE LANE REQUIREMENTS for LANE A
```

```
// Arriving time requirement

//go to change names ONLY
if(timet%2==0){

    int countForRightTurn=0;

    for(int
CheckEnteryGate=0;CheckEnteryGate<A1.length;CheckEnteryGate++)
    {

if((A1[CheckEnteryGate].getSpeed()>=MinLaneA1Speed&A1[CheckEnteryGate].getSp
eed()<=MaxLaneA1Speed) && A1[CheckEnteryGate].getEntering()==0 &&
A1[CheckEnteryGate].getGateArrivalTime()==seconds)
            {

                if(A1[CheckEnteryGate].getTTurn().equals("+"))
                {

                    //System.out.println("The right turn Vehicle");
```

```
                        //double oldA1speed =  A1[CheckEnteryGate].getSpeed();
// This just to show the old speed which rnge between 60-65

A1[CheckEnteryGate].setSpeed(((MaxLaneA1Speed+MinLaneA1Speed)/2.00));
                            A1[CheckEnteryGate].setEntering(1);
                            System.out.println("Right turn vehicle " +
A1[CheckEnteryGate].getVID() + " has entered the intersection through lane [A1] with
speed of " + A1[CheckEnteryGate].getSpeed());
                          //" with time
"+A1[CheckEnteryGate].getGateArrivalTime() +      this shows the entering time


                        }
                        else

                         {

                                 //double oldA1speed =
A1[CheckEnteryGate].getSpeed();    // This just to show the old speed which rnge
between 60-65

A1[CheckEnteryGate].setSpeed(((MaxLaneA1Speed+MinLaneA1Speed)/2.00));
```

```
                                        A1[CheckEnteryGate].setEntering(1);
                                        System.out.println("Vehicle " +
A1[CheckEnteryGate].getVID() + " has entered the intersection through lane [A1] with
speed of " + A1[CheckEnteryGate].getSpeed());
                                        //countForRightTurn++;
                                       //System.out.println("Roundedddd-------
"+(int) Math.ceil((A2[CheckEnteryGate].getExitTime()))+" exit time "
+A1[CheckEnteryGate].getExitTime());
                          }



              }

             for(int
CheckEnteryGate2=0;CheckEnteryGate2<A1.length;CheckEnteryGate2++)
             {
                    //if((int)A1[CheckEnteryGate2].getExitTime()==(int)seconds+1)
                    if(((int) Math.ceil((A1[CheckEnteryGate2].getExitTime())) ==
seconds) & ((A1[CheckEnteryGate2].getEntering()==1))  )
                    {
                    //      System.out.println("second now "
                            A1[CheckEnteryGate2].setEntering(2);
```

```
                                        System.out.println("Vehicle "
+A1[CheckEnteryGate2].getVID()+ " has left the intersection's lane [A1] ");
                                }


                        }


                        //the last curly braclet
                        }



                        for(int
CheckEnteryGateA2=0;CheckEnteryGateA2<A2.length;CheckEnteryGateA2++)
                        {



if((A2[CheckEnteryGateA2].getSpeed()>=MinLaneA2Speed&A2[CheckEnteryGateA2].
getSpeed()<=MaxLaneA2Speed) && A2[CheckEnteryGateA2].getEntering()==0 &&
A2[CheckEnteryGateA2].getGateArrivalTime()==seconds)
                         {

                         if(A2[CheckEnteryGateA2].getTTurn().equals("+"))
                                  {
```

```
                    //System.out.println("The right turn Vehicle");
                //double oldA1speed =  A1[CheckEnteryGate].getSpeed();
// This just to show the old speed which rnge between 60-65

A2[CheckEnteryGateA2].setSpeed(((MaxLaneA2Speed+MinLaneA2Speed)/2.00));
                    A2[CheckEnteryGateA2].setEntering(1);
                    System.out.println("Right turn vehicle " +
A2[CheckEnteryGateA2].getVID() +" has entered the intersection through lane [A2] with
speed of " + A2[CheckEnteryGateA2].getSpeed());



                }

        else
        {

A2[CheckEnteryGateA2].setSpeed(((MaxLaneA2Speed+MinLaneA2Speed)/2.00));
            A2[CheckEnteryGateA2].setEntering(1);
            //System.out.println("Vehicle " + CheckEnteryGateA2 + " has
entered the intersection through lane A2  ");
```

```
                    System.out.println("Vehicle " +
A2[CheckEnteryGateA2].getVID() + " has entered the intersection through lane [A2]
with speed of " + A2[CheckEnteryGateA2].getSpeed());
          }


              }



             for(int
CheckEnteryGateA22=0;CheckEnteryGateA22<A2.length;CheckEnteryGateA22++)
              {
                    if(((int) Math.ceil((A2[CheckEnteryGateA22].getExitTime())) ==
seconds) & ((A2[CheckEnteryGateA22].getEntering()==1)) )
                    {
                          A2[CheckEnteryGateA22].setEntering(2);  //left the
intersection
                          System.out.println("Vehicle "
+A2[CheckEnteryGateA22].getVID()+ " has left the intersection's lane [A2] ");
                    }
              }


              }
```

```
}

if(timet%2==1){

        for(int
CheckEnteryGate1=0;CheckEnteryGate1<B1.length;CheckEnteryGate1++)
        {



if((B1[CheckEnteryGate1].getSpeed()>=MinLaneB1Speed&B1[CheckEnteryGate1].getS
peed()<=MaxLaneB1peed) && B1[CheckEnteryGate1].getEntering()==0 &&
B1[CheckEnteryGate1].getGateArrivalTime()==seconds)
         {
            if(B1[CheckEnteryGate1].getTTurn().equals("+"))
                   {

                         //System.out.println("The right turn Vehicle");
                   //double oldA1speed =  A1[CheckEnteryGate].getSpeed();
// This just to show the old speed which rnge between 60-65

B1[CheckEnteryGate1].setSpeed(((MaxLaneB1peed+MinLaneB1Speed)/2.00));
```

```
B1[CheckEnteryGate1].setEntering(1);
System.out.println("Right turn Vehicle " +
B1[CheckEnteryGate1].getVID() + " has entered the intersection through lane [B1] with
speed of " + B1[CheckEnteryGate1].getSpeed());

}

else
{

B1[CheckEnteryGate1].setSpeed(((MaxLaneB1peed+MinLaneB1Speed)/2.00));
B1[CheckEnteryGate1].setEntering(1);
//System.out.println("Vehicle " + CheckEnteryGateA2 + " has
entered the intersection through lane A2  ");
System.out.println("Vehicle " + B1[CheckEnteryGate1].getVID()
+ " has entered the intersection through lane [B1] with speed of " +
B1[CheckEnteryGate1].getSpeed());
}



}
```

```
for(int
CheckEnteryGate3=0;CheckEnteryGate3<B1.length;CheckEnteryGate3++)
        {
                if(((int) Math.ceil((B1[CheckEnteryGate3].getExitTime())) ==
seconds) & ((B1[CheckEnteryGate3].getEntering()==1)))
                {
                        B1[CheckEnteryGate3].setEntering(2);
                        //System.out.println("Vehicle " +CheckEnteryGate3+ " has
left the intersection's lane B1  ");
                        System.out.println("Vehicle "
+B1[CheckEnteryGate3].getVID()+ " has left the intersection's lane [B1] ");


                }
        }
        }



for(int
CheckEnteryGateB2=0;CheckEnteryGateB2<B2.length;CheckEnteryGateB2++)
        {
```

```
if((B2[CheckEnteryGateB2].getSpeed()>=MinLaneB2Speed&B2[CheckEnteryGateB2].
getSpeed()<=MaxLaneB2peed) && B2[CheckEnteryGateB2].getEntering()==0 &&
B2[CheckEnteryGateB2].getGateArrivalTime()==seconds)
            {
                        if(B2[CheckEnteryGateB2].getTTurn().equals("+"))
                        {

                              //System.out.println("The right turn Vehicle");
                        //double oldA1speed =  A1[CheckEnteryGate].getSpeed();
// This just to show the old speed which rnge between 60-65

B2[CheckEnteryGateB2].setSpeed(((MaxLaneB2peed+MinLaneB2Speed)/2.00));
                              B2[CheckEnteryGateB2].setEntering(1);
                              System.out.println("Right turn Vehicle " +
B2[CheckEnteryGateB2].getVID() + " has entered the intersection through lane [B2]
with speed of " + B2[CheckEnteryGateB2].getSpeed());

                        }
```

```
else
{

B2[CheckEnteryGateB2].setSpeed(((MaxLaneB2peed+MinLaneB2Speed)/2.00));
        B2[CheckEnteryGateB2].setEntering(1);
        //System.out.println("Vehicle " + CheckEnteryGateA2 + " has
entered the intersection through lane A2  ");
        System.out.println("Vehicle " +
B2[CheckEnteryGateB2].getVID() + " has entered the intersection through lane [B2]
with speed of " + B2[CheckEnteryGateB2].getSpeed());
}



}



for(int
CheckEnteryGateB22=0;CheckEnteryGateB22<B2.length;CheckEnteryGateB22++)
{
        if(((int) Math.ceil((B2[CheckEnteryGateB22].getExitTime())) ==
seconds) & ((B2[CheckEnteryGateB22].getEntering()==1)))
        {
```

```
                    B2[CheckEnteryGateB22].setEntering(2);  //left the
intersection
                    System.out.println("Vehicle "
+B2[CheckEnteryGateB22].getVID()+ " has left the intersection's lane [B2] ");
                }
            }

            }



         }



       Thread.sleep(1000);
        timet = timet + 1;
        delay = delay + 1000;

        System.out.println(" ");

      }
      while (delay != 60000);{
      System.out.println("Time's Up!");
```

```
  }

}

public static int randInt(int min, int max)
{

   Random rand = new Random();

   int randomNum = rand.nextInt((max - min) + 1) + min;

   return randomNum;
}

public static double SpeedPerfoot(double a)
{
// a and b are x and y
```

```
        // c is the speed
      double Spf = (((a*5280)/60)/60); // 1 mile equal to 5280 feet. then convert to
minutes , then convert to seconds

      return Spf;
    }



    static class Turn {
            int[] turnAttributes;
            String turnSign;
            public Turn(int[] turnAttributes, String turnSign){
                  this.turnSign = turnSign;
                  this.turnAttributes = turnAttributes;
            }
        }
        //simple class to model results (distance + class)
        static class Result {
            double distance;
            String turnSign;
            public Result(double distance, String turnSign){
                  this.turnSign = turnSign;
```

```
                        this.distance = distance;
                }
        }
        //simple comparator class used to compare results via distances
        static class DistanceComparator implements Comparator<Result> {
                @Override
                public int compare(Result a, Result b) {
                        return a.distance < b.distance ? -1 : a.distance == b.distance
? 0 : 1;
                }
        }

    private static String findMajorityClass(String[] array)
        {

                //add the String array to a HashSet to get unique String values
                Set<String> h = new HashSet<String>(Arrays.asList(array));
                //convert the HashSet back to array
                String[] uniqueValues = h.toArray(new String[0]);
                //counts for unique strings
                int[] counts = new int[uniqueValues.length];
```

```
// loop thru unique strings and count how many times they appear
in origianl array
for (int i = 0; i < uniqueValues.length; i++) {
    for (int j = 0; j < array.length; j++) {
        if(array[j].equals(uniqueValues[i])){
            counts[i]++;
        }
    }
}



for (int i = 0; i < counts.length; i++)
{
  if (i==0)
  {

        System.out.println("Number of points making a right turn :
"+ counts[i]);
  }
  else
  {
     System.out.println("Number of points going ahead : "+ counts[i]);
```

```
    }
}

      int max = counts[0];
      for (int counter = 1; counter < counts.length; counter++) {
            if (counts[counter] > max) {
                  max = counts[counter];
            }
      }


      // how many times max appears
      //we know that max will appear at least once in counts
      //so the value of freq will be 1 at minimum after this loop
      int freq = 0;
      for (int counter = 0; counter < counts.length; counter++) {
            if (counts[counter] == max) {
                  freq++;
            }
      }

      //index of most freq value if we have only one mode
```

```
                int index = -1;
                if(freq==1){
                        for (int counter = 0; counter < counts.length; counter++) {
                                if (counts[counter] == max) {
                                        index = counter;
                                        break;
                                }
                        }
                        //System.out.println("one majority class, index is: "+index);
                        return uniqueValues[index];
                } else{//we have multiple modes
                        int[] ix = new int[freq];//array of indices of modes
                        System.out.println("multiple majority classes: "+freq+"
classes");
                        int ixi = 0;
                        for (int counter = 0; counter < counts.length; counter++) {
                                if (counts[counter] == max) {
                                        ix[ixi] = counter;//save index of each max
count value
                                        ixi++; // increase index of ix array
                                }
                        }
```

```
            for (int counter = 0; counter < ix.length; counter++)
                System.out.println("class index: "+ix[counter]);

            //now choose one at random
            Random generator = new Random();
            //get random number 0 <= rIndex < size of ix
            int rIndex = generator.nextInt(ix.length);
            System.out.println("random index: "+rIndex);
            int nIndex = ix[rIndex];
            //return unique value at that index
            return uniqueValues[nIndex];
        }

    }

private static double meanOfArray(double[] m) {
        double sum = 0.0;
        for (int j = 0; j < m.length; j++){
            sum += m[j];
```

```
            }
            return sum/m.length;
        }

    public static void WriteInstance(int a,int b, int c)
    {
    try {   // this is for monitoring runtime Exception within the block

     String content = (String.valueOf(a)); // content to write into the file
     String content2 = (String.valueOf(b));
     String content3 = (String.valueOf(c));
       File file = new  File("C:\\Users\\User1\\Desktop\\Research\\The
Paper\\instnces.txt"); // here file not created here

       // if file doesnt exists, then create it
       if (!file.exists()) {   // checks whether the file is Exist or not
           file.createNewFile();   // here if file not exist new file created
       }

       FileWriter fw = new FileWriter(file.getAbsoluteFile(), true); // creating
fileWriter object with the file
```

```
BufferedWriter bw = new BufferedWriter(fw); // creating bufferWriter which is used to write the content into the file
// PrintWriter bwn = new PrintWriter(file);   //This to delete the previous content of the file and write a new one

bw.write(content); // write method is used to write the given content into the file
bw.write(" ");
bw.write(content2);
bw.write(" ");
bw.write(content3);
bw.newLine();
bw.close(); // Closes the stream, flushing it first. Once the stream has been closed, further write() or flush() invocations will cause an IOException to be thrown. Closing a previously closed stream has no effect.

System.out.println("Done write Instances");

} catch (IOException e) { // if any exception occurs it will catch
  e.printStackTrace();
}
```

```
}



public static void clearfileInstance()

{

try {   // this is for monitoring runtime Exception within the block



    File file = new  File("C:\\Users\\User1\\Desktop\\Research\\The
Paper\\instnces.txt"); // here file not created here


    // if file doesnt exists, then create it

    if (!file.exists()) {   // checks whether the file is Exist or not

        file.createNewFile();   // here if file not exist new file created

    }


    FileWriter fw = new FileWriter(file.getAbsoluteFile(), true); // creating
fileWriter object with the file

    BufferedWriter bw = new BufferedWriter(fw); // creating bufferWriter which is
used to write the content into the file

   PrintWriter bwn = new PrintWriter(file);   //This to delete the previous content
of the file and write a new one
```

```
// bw.write(" ");
  bw.close(); // Closes the stream, flushing it first. Once the stream has been
closed, further write() or flush() invocations will cause an IOException to be thrown.
Closing a previously closed stream has no effect.

  System.out.println("Done deleting previous Instances");

} catch (IOException e) { // if any exception occurs it will catch
  e.printStackTrace();
}

}

public static void WriteTurn(String a)
{
try {   // this is for monitoring runtime Exception within the block
```

```
String content = (String.valueOf(a)); // content to write into the file
//String content2 = (String.valueOf(b));
// String content3 = (String.valueOf(c));
File file = new  File("C:\\Users\\User1\\Desktop\\Research\\The Paper\\Turn.txt"); // here file not created here

// if file doesnt exists, then create it
if (!file.exists()) {   // checks whether the file is Exist or not
    file.createNewFile();   // here if file not exist new file created
}

FileWriter fw3 = new FileWriter(file.getAbsoluteFile(), true); // creating fileWriter object with the file
BufferedWriter bw3 = new BufferedWriter(fw3); // creating bufferWriter which is used to write the content into the file
// PrintWriter bwn = new PrintWriter(file);   //This to delete the previous content of the file and write a new one
//bw.newLine();
bw3.newLine();
bw3.write(content); // write method is used to write the given content into the file
//bw.write(" ");
```

```
//bw.write(content2);
// bw.write(" ");
//bw.write(content3);
bw3.close(); // Closes the stream, flushing it first. Once the stream has been closed, further write() or flush() invocations will cause an IOException to be thrown. Closing a previously closed stream has no effect.

System.out.println("Done write Turn");

} catch (IOException e) { // if any exception occurs it will catch
e.printStackTrace();
}

}

public static void clearfileTurns()
{
try {   // this is for monitoring runtime Exception within the block
```

```
File file = new  File("C:\\Users\\User1\\Desktop\\Research\\The Paper\\Turns.txt"); // here file not created here

// if file doesnt exists, then create it
if (!file.exists()) {   // checks whether the file is Exist or not
    file.createNewFile();   // here if file not exist new file created
}

FileWriter fw = new FileWriter(file.getAbsoluteFile(), true); // creating fileWriter object with the file
BufferedWriter bw = new BufferedWriter(fw); // creating bufferWriter which is used to write the content into the file
PrintWriter bwn = new PrintWriter(file);   //This to delete the previous content of the file and write a new one



// bw.write(" ");
bw.close(); // Closes the stream, flushing it first. Once the stream has been closed, further write() or flush() invocations will cause an IOException to be thrown. Closing a previously closed stream has no effect.
```

```
  System.out.println("Done deleting previous turns");

} catch (IOException e) { // if any exception occurs it will catch
  e.printStackTrace();
}


}



//b stuff


public static void WriteInstanceB(int a,int b, int c)
{
try {   // this is for monitoring runtime Exception within the block

String content = (String.valueOf(a)); // content to write into the file
String content2 = (String.valueOf(b));
String content3 = (String.valueOf(c));
```

File file = new  File("C:\\Users\\User1\\Desktop\\Research\\The Paper\\instncesB.txt"); // here file not created here

// if file doesnt exists, then create it

if (!file.exists()) {   // checks whether the file is Exist or not

file.createNewFile();   // here if file not exist new file created

}

FileWriter fwb = new FileWriter(file.getAbsoluteFile(), true); // creating fileWriter object with the file

BufferedWriter bwb = new BufferedWriter(fwb); // creating bufferWriter which is used to write the content into the file

// PrintWriter bwn = new PrintWriter(file);   //This to delete the previous content of the file and write a new one

bwb.write(content); // write method is used to write the given content into the file

bwb.write(" ");

bwb.write(content2);

bwb.write(" ");

bwb.write(content3);

bwb.newLine();

bwb.close(); // Closes the stream, flushing it first. Once the stream has been closed, further write() or flush() invocations will cause an IOException to be thrown. Closing a previously closed stream has no effect.

System.out.println("Done write Instances");

} catch (IOException e) { // if any exception occurs it will catch

e.printStackTrace();

}

}

public static void clearfileInstanceB()

{

try { // this is for monitoring runtime Exception within the block

File file = new File("C:\\Users\\User1\\Desktop\\Research\\The Paper\\instncesB.txt"); // here file not created here

```
// if file doesnt exists, then create it
if (!file.exists()) {   // checks whether the file is Exist or not
    file.createNewFile();   // here if file not exist new file created
}

FileWriter fwb2 = new FileWriter(file.getAbsoluteFile(), true); // creating
fileWriter object with the file
BufferedWriter bwb2 = new BufferedWriter(fwb2); // creating bufferWriter
which is used to write the content into the file
PrintWriter bwnb2 = new PrintWriter(file);   //This to delete the previous content
of the file and write a new one



// bw.write(" ");
bwb2.close(); // Closes the stream, flushing it first. Once the stream has been
closed, further write() or flush() invocations will cause an IOException to be thrown.
Closing a previously closed stream has no effect.

System.out.println("Done deleting previous Instances");

} catch (IOException e) { // if any exception occurs it will catch
e.printStackTrace();
```

```
}

}

public static void WriteTurnB(String a)
{
try {   // this is for monitoring runtime Exception within the block

String content = (String.valueOf(a)); // content to write into the file
//String content2 = (String.valueOf(b));
//String content3 = (String.valueOf(c));
 File file = new  File("C:\\Users\\User1\\Desktop\\Research\\The Paper\\TurnB.txt"); // here file not created here

 // if file doesnt exists, then create it
 if (!file.exists()) {   // checks whether the file is Exist or not
     file.createNewFile();   // here if file not exist new file created
 }
```

FileWriter fw4 = new FileWriter(file.getAbsoluteFile(), true); // creating fileWriter object with the file

BufferedWriter bw4 = new BufferedWriter(fw4); // creating bufferWriter which is used to write the content into the file

// PrintWriter bwn = new PrintWriter(file);   //This to delete the previous content of the file and write a new one

//bw.newLine();

bw4.newLine();

bw4.write(content); // write method is used to write the given content into the file

//bw.write(" ");

//bw.write(content2);

// bw.write(" ");

//bw.write(content3);

bw4.close(); // Closes the stream, flushing it first. Once the stream has been closed, further write() or flush() invocations will cause an IOException to be thrown. Closing a previously closed stream has no effect.

System.out.println("Done write Turn");

} catch (IOException e) { // if any exception occurs it will catch

e.printStackTrace();

}

```
}

public static void clearfileTurnsB()
{
try {   // this is for monitoring runtime Exception within the block

 File file = new  File("C:\\Users\\User1\\Desktop\\Research\\The
Paper\\TurnsB.txt"); // here file not created here

 // if file doesnt exists, then create it
 if (!file.exists()) {   // checks whether the file is Exist or not
     file.createNewFile();   // here if file not exist new file created
 }

 FileWriter fw = new FileWriter(file.getAbsoluteFile(), true); // creating fileWriter
object with the file
```

BufferedWriter bw = new BufferedWriter(fw); // creating bufferWriter which is used to write the content into the file

PrintWriter bwn = new PrintWriter(file); //This to delete the previous content of the file and write a new one

// bw.write(" ");

bw.close(); // Closes the stream, flushing it first. Once the stream has been closed, further write() or flush() invocations will cause an IOException to be thrown. Closing a previously closed stream has no effect.

System.out.println("Done deleting previous turns");

} catch (IOException e) { // if any exception occurs it will catch

e.printStackTrace();

}

}

} //End of the whole class

```
package play2;

import java.util.ArrayList;
import java.util.HashMap;
import java.util.Map;

public class Car2 {

    private double speed;  // the speed of the car
    //private int feetPerSecond; // convert from mile/hour to feet/second
    private int TheX;
    private int TheY;
    private int TheEntering; // 0 not entered     1 entered    2 left
    private int VehicleId;
    private double TheArrival;
    private int Direction; //01 going to the A1  ,02 going to the A2, 11 going to
the B1 , 12 going to the B2
    private int RightTurn; //Right turn number  -1 if there is no right turn
    private double TheStayingTime;
```

```
    private double TheExitingTime;


    private int TDay;
    private int THour;
    private int TEvent;
    private String TTurn;


    public double setSpeed(double s) {


        this.speed=s;
         return s;
    }


    public double getSpeed() {
        return this.speed;
}


    public void setTheX(int x){
        this.TheX=x;
    }
    public int getTheX(){
        return this.TheX;
```

```
}
public void setTheY(int y){
	this.TheY=y;
}
public int getTheY(){
	return this.TheY;
}

public void setDirection(int d){
	this.Direction=d;
}
public int getDirection(){
	return this.Direction;
}

public void setTurn(int t){
	this.RightTurn=t;
}
public int getTurn(){
	return this.RightTurn;
}
```

```
public void setVID(int vid){
        this.VehicleId=vid;
}
public int getVID(){
        return this.VehicleId;
}



public void setDay(int day){
        this.TDay=day;
}
public int getDay(){
        return this.TDay;
}



public void setHour(int hour){
        this.THour=hour;
}
public int getHour(){
        return this.THour;
```

```
}



public void setEvent(int event){
	this.TEvent=event;
}
public int getEvent(){
	return this.TEvent;
}

public void setTTurn(String tturn){
	this.TTurn=tturn;
}
public String getTTurn(){
	return this.TTurn;
}



public void setEntering(int en){
	this.TheEntering=en;
}
public int getEntering(){
```

```
        return this.TheEntering;

}

public void setGateArrivalTime(double ar){
        this.TheArrival=ar;
}
public double getGateArrivalTime(){
        return this.TheArrival;

}

public void setStayingTime(double st){
        this.TheStayingTime=st;
}
public double getStayingTime(){
        return this.TheStayingTime;
}

public void setExitTime(double ex){
        this.TheExitingTime=ex;
}
```

```
        public double getExitTime(){

                return this.TheExitingTime;



        }

}
```